\documentclass[fleqn,usenatbib]{mnras}

\usepackage{graphicx}	% Including figure files
\usepackage{amsmath}	% Advanced maths commands
\usepackage{amssymb}	% Extra maths symbols
\usepackage{url}
\usepackage{ragged2e}
\usepackage{wrapfig}
\usepackage{textcomp}
\usepackage{hyperref}
\usepackage{setspace}
\usepackage{fancyhdr}
\usepackage{pdfpages}
\usepackage{float}
\usepackage{epstopdf}
\usepackage[toc, page]{appendix}

\title[Dark energy with oscillatory tracking potential]{Dark energy with oscillatory tracking potential: Observational Constraints and Perturbative effects}

\author[Albin Joseph and Rajib Saha]{
	Albin Joseph\thanks{E-mail: albinje@iiserb.ac.in}
	and Rajib Saha\thanks{E-mail: rajib@iiserb.ac.in}
	\\
$^{1}$Department of Physics, Indian Institute of Science Education and Research (IISER) Bhopal, 462066, India\\
}

\date{Accepted XXX. Received YYY; in original form ZZZ}

\pubyear{2021}

\begin{document}
\label{firstpage}
\pagerange{\pageref{firstpage}--\pageref{lastpage}}
\maketitle

\begin{abstract}
The cosmological models exhibiting tracker properties have great significance in the context of dark energy as they can reach the present value of dark energy density from a wide range of initial conditions, thereby alleviating both the fine-tuning and the cosmic coincidence problem. The  $\alpha$-attractors, which are originally discussed in the context of inflation, can exhibit the properties of dark energy as they can behave like cosmological trackers at early times and show the late time behaviour of a cosmological constant. In the present paper, we study the Oscillatory Tracker Model (OTM), which belongs to the family of $\alpha$-attractor dark energy models.  Using the current observational data sets like Cosmic Microwave Background (CMB), Baryon Acoustic Oscillation (BAO) and type 1a supernova data (Pantheon compilation), we constrain the parameters of the model and estimate both the mean and best-fit values. Although the oscillatory tracker model contains a larger set of parameters than the usual LCDM model, the common set of parameters of both agree within $1\, \sigma$ error limits. Our observations using both high redshift and low redshift data supports Hubble parameter value $H_0 = 67.4$ \,Kms$^{-1}$Mpc$^{-1}$. We study the effect of the OTM on the CMB temperature and polarization power spectra, matter power spectrum and $f \sigma_8$. Our analysis of the CMB power spectrum and matter power spectrum suggests that the oscillatory tracker dark energy model has noticeable differences from usual LCDM predictions. Yet, in most cases, the agreement is very close.
\end{abstract}


%%%%%%%%%%%%%%%%%%%%%%%%%%%%%%%%%%%%%%%%%%%%%%%%%%

%%%%%%%%%%%%%%%%% BODY OF PAPER %%%%%%%%%%%%%%%%%%

%-------------------------
%-------------------------
\section{Introduction}
\label{sec:intro}
%-------------------------
%-------------------------

The accelerated expansion of the universe, which was first discovered through the observations of type 1a supernovae~\citep{Riess_1998, Perlmutter_1999} serves as a paradigm shift in our understanding of cosmology. This hypothesis was strongly supported by the data from other observations like CMB anisotropies~\citep{2020} or the large scale structure~\citep{Eisenstein_2005,Blake_2012,Parkinson_2012,Kazin_2014,Beutler_2015}. The current accelerated expansion is attributed to the so-called 'dark energy', and it provides a dominant contribution to the present total energy density of the universe. The contribution of the matter content of the universe is represented by the energy-momentum tensor on the right-hand side of the Einstein equation, whereas the left-hand side is represented by pure geometry. The accelerated expansion of the universe can be obtained either by supplementing the energy-momentum tensor by an exotic form of matter such as cosmological constant or scalar field and also by modifying the geometry itself. 

Presently the most accepted cosmological model is the LCDM model, which consists of a cosmological constant $\Lambda$, cold dark matter, baryons, photons and neutrinos. Though it can explain many observed cosmological phenomena, it has several theoretical inconsistencies like the fine-tuning~\citep{Sahni_2002} and the cosmic coincidence problems~\citep{Steinhardt:2003st, Velten_2014}. This motivates cosmologists to look forward to alternate models for the explanation of cosmological dilemmas. So instead of considering a cosmological constant, dark energy is introduced as a dynamical phenomenon. Among them, the models such as Quintessence~\citep{Linder_2007,Tsujikawa_2013,Chiba_2013,Durrive_2018}, Phantoms~\citep{Caldwell_2002, Caldwell_2003,Nojiri_2005,PhysRevD.98.043519}, K-essence~\citep{Chiba_2000,Armendariz_Picon_2000,Armendariz_Picon_2001}, Tachyon~\citep{Padmanabhan_2002,Bagla_2003,Abramo_2003,Aguirregabiria_2004,Guo_2004,Copeland_2005} and Dilatonic Dark Energy~\citep{Piazza_2004,Damour_2002} consider a scalar field as responsible for the dynamics of the dark energy. In an alternative approach, the accelerated expansion of the universe can be obtained from the geometrical modifications which can arise from quantum effects such as higher curvature corrections to the Einstein Hilbert action~\citep{lobo2008dark,Tsujikawa_2011,Li_2011,Clifton_2012,Dimitrijevic_2013,Brax_2015,Joyce_2016}. 

As discussed above, the dark energy models are very successful in explaining the current accelerated expansion of the universe. On the other hand, the primordial inflationary models~\citep{PhysRevD.23.347} play a significant role in explaining other observed phenomena like the origin of CMB anisotropies~\citep{Abazajian_2015} and the formation of the large scale structure~\citep{L_Huillier_2018}. 
The origin of early and late inflation still remains a theoretical puzzle that motivates the theorists to simultaneously explain both inflationary phases by invoking scalar fields. The scalar fields play a fundamental role in cosmology as they are simple, yet a natural candidate for the accelerated expansion of the universe. Presently there are wide variety of inflationary models~\citep{PhysRevD.23.347, Linde:1981mu, Linde:1983gd} that have been proposed, and among them, the cosmological attractor models were discovered recently~\citep{Kallosh_2013}. These cosmological attractor models belong to a wide class of cosmological models which incorporate the conformal attractors~\citep{Kallosh_2013}, alpha attractors~\citep{Kaiser_2014,Kallosh_2013alpha,Kallosh_2014,Miranda_2017,Shahalam_2018}, and also include scalar field cosmological models such as the Starobinsky model~\citep{Starobinsky:1980te,Mukhanov:1981xt,Starobinsky:1983zz,Whitt:1984pd,Kofman:1985aw}, the chaotic inflation in supergravity (GL model)~\citep{Goncharov:1984jlb,Goncharov:1983mw,Linde_2015}, the Higgs inflation~\citep{PhysRevD.40.1753,Bezrukov_2008,Cervantes_Cota_1995,PhysRevD.82.045003,Linde_2011} and the axion monodromy inflation~\citep{PhysRevD.78.106003,McAllister_2010,Conlon_2012,Flauger_2010,Brown_2016}. The conformal attractor models predicts that for a large number of e-folds $N$, the spectral index and tensor-to-scalar ratio are given by $n_s = 1-2/N ;\, r = 12/N^2$ whereas for alpha attractors, the slow-roll parameters are given by $n_s = 1-2/N ;\, r = 12 \alpha/N^2$ for small $\alpha$ and $n_s = 1-2/N ;\, r = 12\alpha/(N(N+3\alpha/2))$ for large $\alpha$. Although these models have different origins, they provide very similar cosmological predictions with WMAP~\citep{Hinshaw_2013} and the recently released Planck data~\citep{2020}. These models can be used not only for inflation but also for late-time cosmic acceleration. In the context of dark energy, the cosmological models with tracker properties have gained great attention as the scalar field can reach the present value of dark energy density from a wide range of initial conditions. Thus a scalar field rolling down a slowly varying potential not only gives rise to the current accelerated expansion but also alleviates the cosmic coincidence problem.

In the present work, we focus on a specific $\alpha$-attractor dark energy model -  the  Oscillatory Tracker Model (OTM), which was initially proposed in~\citep{Bag_2018}. The OTM, with its tracker properties, can alleviate the cosmic coincidence problem and moreover, it is very much favoured over various other $\alpha$-attractor dark energy models~\citep{Cede_o_2019}. Here we focus on testing the OTM against different cosmological observations like type 1a supernovae, BAO, and CMB. We also study the effect of the OTM on the CMB temperature and polarization power spectra, matter power spectrum and $f\sigma_s$ with respect to the background LCDM model. 

This paper is structured as follows. In section~\ref{Sec:Model}, we present the basics of the $\alpha$-attractor dark energy model and the oscillatory tracker dark energy model. The background equations which determine the dynamics of the dark energy model is described in section~\ref{Sec:Background}. After considering linear perturbation around the Friedmann-Lemaitre-Robertson-Walker (FLRW) background in section~\ref{Sec:linear}, we move to section~\ref{Sec:Methods} where we describe the datasets and methodology used to constrain the parameters of the model, and we also quantify our results through 2D posterior, best-fit and mean values. The section~\ref{Sec:effects} is dedicated to the study of the effect of the oscillatory tracker dark energy model on CMB temperature and polarization power spectra, matter power spectrum and $f \sigma_8$.  Finally, in section~\ref{Sec:Conclusions} we discuss and conclude upon our results.

%-------------------------
%-------------------------
\section{Oscillatory Tracker Dark Energy Model} \label{Sec:Model}
%-------------------------
%-------------------------

\begin{figure}
	\centering
	\includegraphics[width=0.5\textwidth]{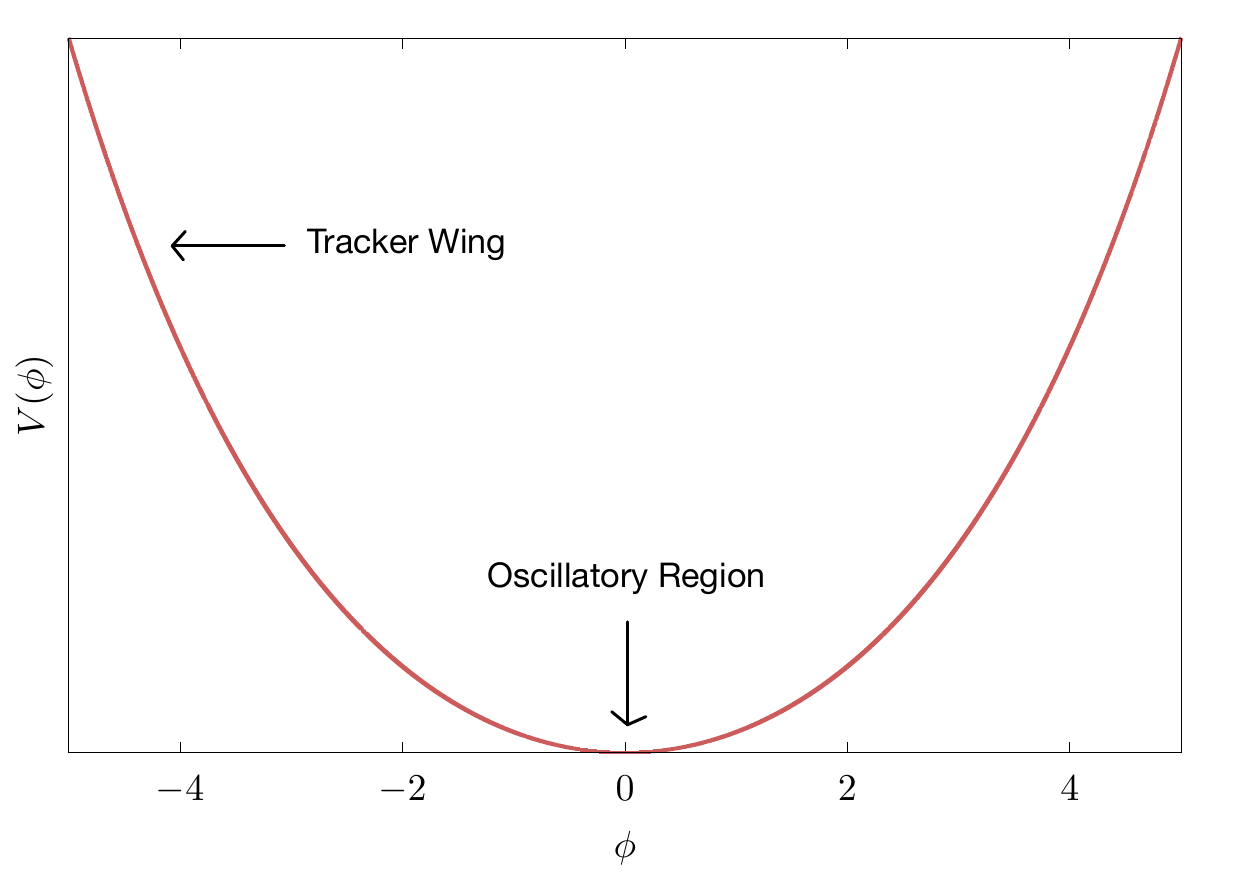}
	\caption{\small A schematic representation of the potential energy density for the oscillatory tracker dark energy model. \normalsize
	}
	\label{fig:potential}
\end{figure}

The $\alpha$-attractors have been gaining attention in the context of dark energy due to their possibility of linking both the inflationary and the present accelerated expansion of the universe. Their predictions in the inflationary paradigm are in good agreement with the latest cosmological observations~\citep{Akrami_2018}, and as quintessence models, they can also produce late time accelerated expansion compatible with the present measurements~\citep{Garc_a_Garc_a_2018}. In this article, we focus on the minimally coupled $\alpha$-attractor dark energy model with the Lagrangian density in the Einstein frame represented by,
\begin{equation}
	\mathcal{L}=\sqrt{-g}\left[ \frac{1}{2} M_p^2R-\frac{\alpha }{\left( 1-\frac{\phi ^{2}}{6}%
		\right) ^{2}}\frac{\left( \partial \phi \right) ^{2}}{2}-\alpha
	f^{2}\left( \frac{\phi }{\sqrt{6}}\right) \right],
	\label{eq:lag}
\end{equation}
where $M_p$ is the Planck mass, $\alpha$ is a parameter and $\alpha f^2$ is the potential function dependent on the field $\phi$ which is measured in $M_p$ units. Here the kinetic term is not canonical, but can be made canonical by a field redefinition $\varphi =\sqrt{
	6\alpha }\tanh ^{-1}\left( \frac{\phi }{\sqrt{6}}\right) $. Now the Lagrangian density can be written as,
%\vspace{-0.5cm}
\begin{equation}
	\mathcal{L}=\sqrt{-g}\left[ \frac{1}{2} M_p^2R-\frac{\left( \partial \varphi \right) ^{2}}{2}-\alpha
	f^{2}\left( x\right) \right],
	\label{eq:newlag}
\end{equation}
where $x=\tanh\left(\frac{\varphi}{\sqrt{6 \alpha}}\right)$. This implies one can write the scalar field potential as  $V\left( \varphi \right)=\alpha f^{2}\left( \tanh \left( 
\frac{\varphi }{\sqrt{6\alpha }}\right) \right)$. Amoung the various realizations of $\alpha$-attractor dark energy models studied in~\citep{Bag_2018}, we focused our attention on the oscillatory tracker model given by,
%\vspace{-0.5cm}
\begin{equation}
	V(\varphi) = \alpha c^2 \cosh \left(\frac{\varphi}{\sqrt{6 \alpha}}\right),
	\label{eq:potential}
\end{equation}
where the constants $\alpha$ and $c$ are the free parameters. The schematic representation of the oscillatory dark energy model is shown in figure~\ref{fig:potential}. 	 For large values of $\frac{\mid \varphi \mid}{\sqrt{6 \alpha}} >>1$, the oscillatory tracker potential has the asymptotic form,
	$V(\varphi)\simeq \alpha c^2 \exp({\frac{\varphi}{\sqrt{6 \alpha}}})$. So initially, the oscillatory tracker potential behaves like an exponential potential which exibits a very large initial basin of attraction and has been extensively studied in~\citep{PhysRevD.37.3406,Ferreira:1997au,Ferreira:1997hj,Barreiro:1999zs}. During this period the OTM tracks the background density fields. Due to this exponential tracker asymptote, the oscillatory tracker model can avoid the fine tuning problem which affects many models of dark energy. Moreover, it is also worth to note that in OTM, the initial density values of scalar field covering a range of more than 40 orders of magnitude at $z=10^{12}$ can converge onto the attractor scaling solution~\citep{Bag_2018}. This range substantially increases if we set our initial conditions at earlier times. Once the scalar field rolls down the exponential potential (tracker wing), it starts oscillating at the minimum of the potential (oscillatory region). For small values $\frac{\mid \varphi \mid}{\sqrt{6 \alpha}} <<1$, the potential has the limiting form $V(\varphi) \simeq \alpha c^2 (1+\frac{1}{2}(\frac{\varphi}{\sqrt{6 \alpha}})^2)$. As a result, at the late time, the oscillatory tracker potential behaves like a cosmological constant $\alpha c^2$. However, because of the presence of $\varphi ^2$ term, the equation of state parameter of the scalar field approaches $\omega_{\phi}\simeq-1$ via small oscillations.
	Thus due to the presence of the exponential tracker asymptote, the oscillatory tracker model has a very large initial basin of attraction, trajectories from which get funneled into the late time attractor $\omega_{\phi}\simeq-1$. For the detailed theoretical study of the properties of this potential, we refer to the work in~\citep{Bag_2018}.

%-------------------------
%-------------------------
\section{Background Evolution} \label{Sec:Background}
%-------------------------
%-------------------------

In the spatially flat homogeneous and isotropic model of the universe, the space-time interval $ds$ between two events in a global comoving cartesian coordinate system is described by,
\begin{equation}
	ds^2 =  a^2(\tau)(-d\tau^2 + \delta_{ij}dx^idy^j ),
	%    ds^2 = -dt^2 + a^2(t)(dx^2 + dy^2 + dz^2)
\end{equation}
where $\tau$ represents the conformal time which is related to the comic time $t$ as $a^2d\tau^2 = dt^2$ and $a(\tau)$ is the scale factor of expansion which satisfies the Friedmann equation,
\small
\begin{equation}
	\mathcal H^2 \equiv \left(\frac{a^{'}}{a}\right)^2 = \frac{8\pi G}{3} a^2 \rho_{tot} = \frac{8\pi G}{3} a^2 \left(\rho_{\gamma} + \rho_{\nu}+\rho_b + \rho_c +\rho_{\varphi}\right),
\end{equation}
\normalsize
\hspace{-0.1cm}where $\rho_{tot}$  is the total background energy density of all species namely photons ($\gamma$), neutrinos ($\nu$), baryons ($b$), cold dark matter ($c$) and a scalar field ($\varphi$) with a potential $V(\varphi)$ acting as dark energy. In this article, we use $c=1$, reduced Planck mass $M_p=1$ and the prime ($'$) represents derivative with respect to the conformal time $\tau$. The equation of state parameter is given by $\omega_i=\frac{p_i}{\rho_i}$ where $p_i$ corresponds to the pressure of each species. For the photons and neutrinos, $\omega_{\gamma} = \omega_{\nu} =1/3$, whereas for baryons and cold dark matter, $\omega_b = \omega_c =0$. The background energy density and pressure for the scalar field are,
\begin{equation}
	\rho_{\varphi} = \frac{1}{2a^2}\varphi^{'2}+V(\varphi),
	\label{eq:scden}
\end{equation}
\begin{equation}
	p_{\varphi} = \frac{1}{2a^2}\varphi^{'2}-V(\varphi).
	\label{eq:scpre}
\end{equation}
Using Eq.~\ref{eq:scden} and Eq.~\ref{eq:scpre} the equation of state parameter of the scalar field reads,
\begin{equation}
	\omega_{\varphi} = \frac{p_{\varphi}}{\rho_{\varphi}} = \frac{\frac{1}{2a^2}\varphi^{'2}-V(\varphi)}{\frac{1}{2a^2}\varphi^{'2}-V(\varphi)}.
\end{equation}
The background Klein-Gordon equation can be obtained as a consequence of the Bianchi identities as
\begin{equation}
	\varphi^{''} + 2\,\mathcal{H}\,\varphi^{'} + a^{2}\frac{dV}{d\varphi}=0.
\end{equation}

\begin{figure}
	\centering
	\includegraphics[width=0.5\textwidth]{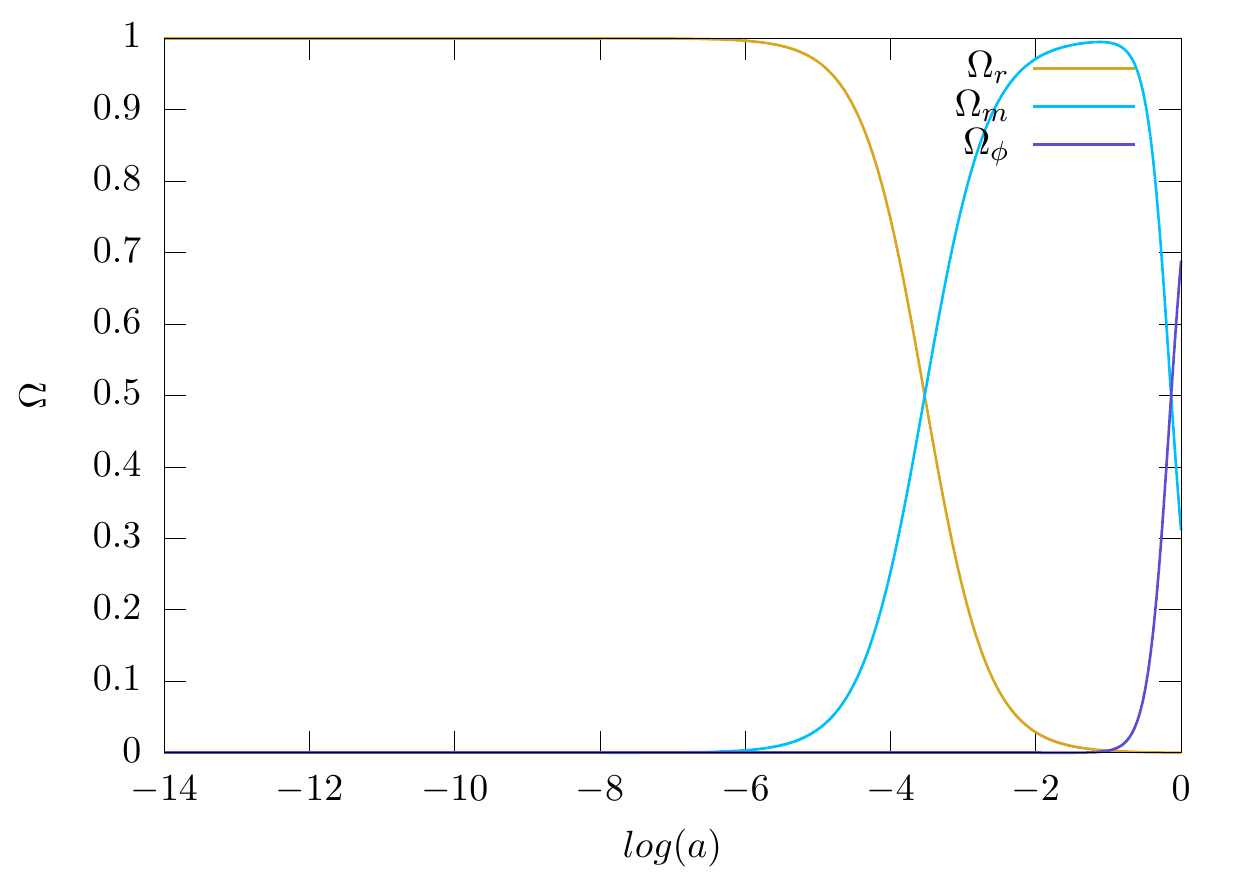}
	\caption{\small
		Figure showing the variation of density parameter $\Omega$ as a function of scale factor, $a$. Here the radiation density $\Omega_r$ and matter density $\Omega_m$ are plotted along with scalar field  density $\Omega_{\phi}$.\normalsize
		%%   
		%  Here the radiation density $\Omega_r$ which incorporates photons ($\gamma$) and neutrinos ($\nu$) and %matter density which incorporating both baryons ($b$) and cold dark matter ($c$) are plotted along with %scalar field ($\phi$).
	}
	\label{fig:density}
\end{figure}

The evolution of the energy density parameters of radiation $\Omega_r$ where ($r = \gamma +\nu$), matter $\Omega_m$ where ($m = b +c$) and scalar field $\Omega_{\phi}$ with the logarithmic of scale factor is shown in figure~\ref{fig:density}. The scalar field tracks both the background fields ($r\,,m$) at the early epochs and in the recent era it dominates over the background fields.

%-------------------------
%-------------------------
\section{Linear Perturbations} \label{Sec:linear}
%-------------------------
%-------------------------

In order to study the observational effects of the $\alpha$-attractor dark energy models on the CMB and large scale structure, we need to consider linear perturbations around the FLRW background. The scalar perturbation of the FLRW metric takes the form~\citep{V_liviita_2008},
\begin{equation}
	\begin{split}
		ds^2 = a^2(\tau) \Big[-(1+2\Phi)d\tau^2 + 2 \partial_iB\,d\tau\,dx^i +\\ \big( (1-2\psi)\delta_{ij} + 2\partial_i\,\partial_jE \big)\,dx^i\,dx^j \Big],
	\end{split}
\end{equation}
where $\Phi, \psi, B, E$ are gauge-dependent functions of both space and time. In synchronous gauge $\Phi\,=\,B=0,\,\psi\,= \eta$ and $k^2E =-h/2-3\eta$, where $\eta$ and $h$ are the synchronous gauge fields defined in the Fourier space and $k$ is the wave number~\citep{1995ApJ...455....7M}. In Fourier space, the perturbation equations in the matter sector reads as,
\begin{eqnarray}
	\delta_{i}^\prime+ k \,v_{i} +\frac{h^\prime}{2} &=& 0, \\ 
	v_{i}^\prime+\mathcal H v_{i}&=& 0, 
\end{eqnarray}
where $\delta_{i} = \delta \rho_{i}/\rho_{i}$ is the density contrast and $v_{i}$ is the peculiar velocity of $i$-th $(i= b,c)$ fluid. Assuming there is no momentum transfer in CDM frame, we set $v_c$ to zero. For the details of these sets of equation, we refer to the works of~\citep{1995ApJ...455....7M,1984PThPS..78....1K,1992PhR...215..203M,PhysRevD.67.063516}.
The linearized scalar field equation in the Fourier space with wave number $k$ is given by,
\begin{equation} \label{eq:kg2}
	\delta \varphi^{\prime \prime} + 2 \mathcal H \delta \varphi^{\prime} + k^{2}\delta \varphi +a^{2}~\frac{d^{2} V}{d \varphi^{2}} \delta \varphi + \frac{1}{2} \varphi^{\prime} h^{\prime}= 0,
\end{equation}
\hspace{-0.1cm}where $V(\varphi)$ is the potential energy density corresponding to the oscillatory model given in Eq.~\ref{eq:potential}. The perturbation in the energy density $\delta\rho_{\phi}$ and pressure $\delta p_{\phi}$ are,
\begin{eqnarray}
	\delta \rho_{\varphi} &=& -\delta T_{0 (\varphi)}^{0}~=~ \frac{\varphi^{\prime} \delta \varphi^{\prime}}{a^{2}}+\delta \varphi \frac{d V}{d \varphi} , \label{eq:pe2}\\
	\delta T_{0(\varphi)}^{j} &=& - \frac{ i k_{j}\, \varphi^{\prime}\, \delta \varphi}{a^{2}},  \label{eq:pv2}\\
	\delta p_{\phi} \delta^{i}_{j}&=& \delta T_{j(\varphi)}^{i}~=~ \Bigg(-\frac{\varphi^{\prime} \delta \varphi^{\prime}}{a^{2}}-\delta \varphi \frac{d V}{d \varphi}\Bigg)\delta^{i}_{j}\label{eq:pp2},
\end{eqnarray}
here $\delta T_{j(\varphi)}^{i}$ is the perturbed stress-energy tensor of the scalar field. For an adiabatically expanding universe, the square of sound speed is $c_{s,\varphi}^2 = p_{\phi}^{\prime}/\rho_{\varphi}^{\prime}$. We implemented the above equations in CLASS~\citep{Blas_2011,lesgourgues2011cosmic} with adiabatic initial conditions in order to compute the CMB temperature and polarization power spectra and matter power spectrum.

%-------------------------
%-------------------------
\section{Observational Constraints} \label{Sec:Methods}
%-------------------------
%-------------------------

In this section, we compare the oscillatory tracker model with recent observational data. The motivation of this section is to obtain the best-fit and mean values of the cosmological parameters when the oscillatory tracker model is taken into account. In section~\ref{Sec:cmbdata}, we explain the datasets used to constrain the parameters of the model. The analysis of these datasets will help us to accurately determine the implications of the oscillatory tracker model on the CMB power spectra, matter power spectrum and the $f \sigma_8$, which we plan to study in the next section. 
\begin{figure*}
	\centering 
	\includegraphics[width=1\textwidth]{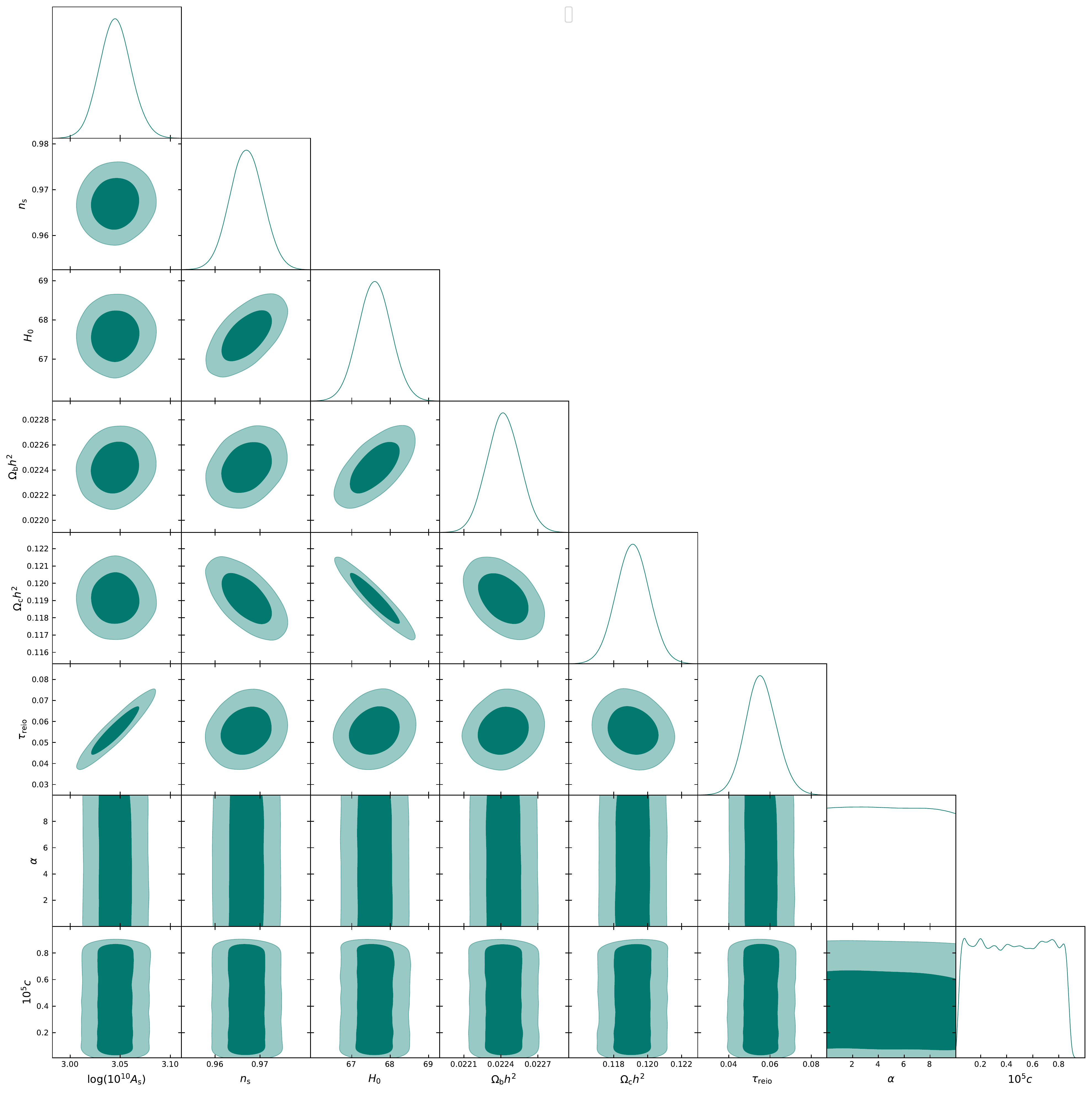}
	\caption{\small The posterior distribution for the model parameters of the OTM. The contours show $68.3\%$ and $95.5\%$ confidence regions. A noticeable correlation exists between the parameter pairs ($H_0$, $\Omega_ch^2$) and ($\tau_{reio}$, $log(10^{10}A_s$). The other parameter pairs ($\Omega_bh^2$, $H_0$), ($\Omega_ch^2$, $n_s$) and ($\Omega_ch^2$, $\Omega_bh^2$) also exibit minor degree of correlations. The quantitative results are summarised in table~\ref{tab:best_fit}. \normalsize
	}
	\label{fig:posteriors}
\end{figure*}
\begin{table}
	\centering
	\begin{tabular}{lc} 
		\hline 
		Parameter   &   Prior \\ 
		\hline 
		\hline
		$\Omega_\mathrm{b} h^2$
		&  $[0.005, 0.1]$
		\\
		$\Omega_\mathrm{c} h^2$
		& 	$[0.001,0.99]$
		\\
		$\tau_{reio}$
		& $[0.01,0.8]$
		\\
		$H_0$
		& $[20, 100]$
		\\
		$n_s$
		& $[0.8, 1.2]$
		\\    	 
		$log(10^{10}A_s)$
		& $[1.61, 3.91]$
		\\
		$\varphi_{\rm i} $
		& 	$10$
		\\
		$10^5c$
		& $[0.01,1.0]$
		\\
		$\alpha$
		& $[0.0,10.0]$
		\\
		\hline 
	\end{tabular} 
	\caption{\small The range of priors used for MCMC analysis. \normalsize
		\label{tab:priors}
	}
\end{table}
\subsection{Data Sets} \label{Sec:cmbdata}

\textit{\textbf{CMB:-}} The data from CMB is very powerful in constraining the dark energy models. Here we used the latest CMB data from the Planck 2018 final data release~\citep{2020A&A...641A...5P}. The CMB likelihood consists of the low-$\ell$ temperature likelihood, $C_{\ell}^{TT}$, low-$\ell$ polarization likelihood, $C_{\ell}^{EE}$, high-$\ell$ temperature-polarization likelihood, $C_{\ell}^{TE}$ and high-$\ell$ combined TT, TE and EE likelihood. The low-$\ell$ CMB likelihood covers the multipole range $2\leq \ell\leq29$ whereas the high-$\ell$ likelihood spans the multipole range $\ell \geq 30$.
\\
\\
%\subsection{Type 1a Supernovae } \label{Sec:novadata}
\textit{\textbf{Type 1a supernovae:-}} The type 1a supernovae, considered standard candles, is an ideal probe for studying cosmological expansion. In this article, we consider the type 1a supernovae data from the Pantheon compilation~\citep{2018ApJ...859..101S}. This consists of 1048 type 1a supernovae data points distributed in the redshift interval $0.01<z<2.26$.
\\
\\
%\subsection{BAO} \label{Sec:baodata}
\textit{\textbf{BAO:-}} The fluctuations in the photon-baryon fluid in the early universe leave their imprints as acoustic peaks in the CMB angular power spectrum. These anisotropies in the baryon acoustic oscillations provide tighter constraints on the cosmological parameters. Here we consider the BAO data from different astronomical surveys 6dFGS~\citep{2011MNRAS.416.3017B}, BOSS DR12~\citep{2017MNRAS.470.2617A} and SDSS main galaxy sample~\citep{2015MNRAS.449..835R}.
\\
\subsection{Methodology and Posterior Analysis} \label{Sec:baodata}
The posterior distribution for the oscillatory tracker model is obtained by sampling the parameter space with a Markov Chain Monte Carlo method (MCMC). In particular, to obtain the best-fit and the mean values of the cosmological parameters, we make use of MCMC simulator Cobaya~\citep{Torrado_2021} and a modified version of the CLASS~\citep{Blas_2011,lesgourgues2011cosmic}. For the statistical analysis of the MCMC results, we make use of the publically available GetDist~\citep{lewis2019getdist} software package. We sample the posterior parameter distribution until the Gelman-Rubin convergence statistic~\citep{Gelman:1992zz} satisfies $R-1 < 0.01$. The parameter space for constraining the oscillatory tracker dark energy model is,

$P \equiv [{{\Omega_bh^2,\Omega_ch^2, H_0, n_s, \tau_{reio},ln(10^{10}A_s), \alpha, 10^5c}}]$,\\
where $\Omega_bh^2$ is the baryon density, $\Omega_ch^2$ is the cold dark matter density, $\tau_{reio}$ is the optical depth to reionization, $H_0$ is the Hubble constant, $A_s$ is the scalar primordial power spectrum amplitude, $n_s$ is the scalar spectral index and $\alpha$ and $c$ are the free model parameters. Moreover, the initial condition for the scalar field velocity $\dot{\varphi_i}$ is set to zero, and since, the same results are obtained for the different initial values of the scalar field $\varphi_i$, we fixed the scalar field initial value $\varphi_i$ to be 10. The parameter space, $P$ for the oscillatory model is explored for the flat prior ranges given in table~\ref{tab:priors}.

\begin{table}
	\addtolength{\tabcolsep}{1pt}
	\renewcommand{\arraystretch}{1.4}
	\centering
	\begin{tabular}{|l|c|c|}
		\hline 
		%	\text{Parameter} &  \text{Best-fit  }   & \text{Best-fit  }   & \text{Best-fit  } \\
		\text{Parameter} &  \text{Best-fit $\pm$ }   & \text{Mean $\pm$} \\
		\text{} &  \text{ 95.5\% limits}   & \text{ 95.5\% limits}\\
		\hline
		{$\Omega_\mathrm{b} h^2$}       & $0.02243^{+0.00056}_{-0.00045}$      & $0.02242 \pm 0.00026$    \\
		{$\Omega_\mathrm{c} h^2$}       & $0.1192^{+0.0038}_{-0.0040}$      & $0.1191 \pm 0.0019$   \\
		{$n_\mathrm{s}$}       & $0.9672^{+0.014}_{-0.015}$      & $0.9670 \pm 0.0072$ \\
		{$H_0$}       & $67.474^{+1.86}_{-1.60}$      & $67.5989^{+0.8498}_{-0.8401}$ \\
		{$\ln(10^{10} A_\mathrm{s})$}  & $3.0417^{+0.075}_{-0.057}$         & $3.0452^{+0.0324}_{-0.0308}$  \\
		{$\tau_\mathrm{reio}$}       & $0.05247^{+0.036}_{-0.028}$      & $0.0557^{+0.0157}_{-0.0148}$  \\
		{$\alpha$}       & $5.5983^{+4.40}_{-5.59}$      & $4.972^{+4.972}_{-5.028}$  \\
		{$10^{5} c$}       & $0.52199^{+0.390}_{-0.496}$      & $0.4546^{+0.4006}_{-0.4052}$ \\
		\hline 
	\end{tabular}
	\caption{\small Best-fit and mean values with $95.5\%$ intervals for the parameters of the oscillatory tracker model. Both best-fit and mean values of the parameters expect $\alpha$ and $10^5 c$ agree very well with each other. $H_0$ units are km\,s$^{-1}$\,Mpc$^{-1}$ and Mpc$^{-2}$ for $c^2$. \normalsize}
	\label{tab:best_fit}
\end{table}

\begin{table}
	\addtolength{\tabcolsep}{1pt}
	\renewcommand{\arraystretch}{1.4}
	\centering
	\begin{tabular}{|l|c|}
		\hline 
		%	\text{Parameter} &  \text{Best-fit  }   & \text{Best-fit  }   & \text{Best-fit  } \\
		\text{Parameter} &  \text{ Mean $\pm$ 68\% limits}    \\
		\text{} &  \text{(Planck 2018)} \\
		\hline
		{$\Omega_\mathrm{b} h^2$}       & $0.02236 \pm 0.00015$         \\
		{$\Omega_\mathrm{c} h^2$}       & $0.1201 \pm 0.0014$       \\
		{$n_\mathrm{s}$}       & $0.9649 \pm 0.0044$     \\
		{$H_0$}       & $67.27 \pm 0.60$      \\
		{$\ln(10^{10} A_\mathrm{s})$}  & $3.045 \pm 0.016$       \\
		{$\tau_\mathrm{reio}$}       & $0.0544^{+0.0070}_{-0.0081}$      \\
		%	{$\alpha$}       & $5.5983^{+4.40}_{-5.59}$      \\
		%	{$10^{5} c$}       & $0.52199^{+0.390}_{-0.496}$       \\
		\hline 
	\end{tabular}
	\caption{\small The mean values with $68\%$ intervals for the parameters of the LCDM model from Planck 2018~\citep{2020}. The LCDM values agree excellently with the corresponding parameter values of OTM case shown in table~\ref{tab:best_fit}. \normalsize} 	\label{tab:best_fit_planck}
\end{table}

The posterior distributions for the parameters of the oscillatory tracker model are shown in figure~\ref{fig:posteriors}. The contours show $1 \sigma$ region of $68\%$ confidence level and $2 \sigma$ region of $95\%$ confidence level with the darker colour signifying the more probable results. From the quantitative results summarized in table~\ref{tab:best_fit}, it is interesting to note that the standard cosmological parameters of the oscillatory tracker model are in good agreement with the LCDM Planck 2018 results~\citep{2020} given in table~\ref{tab:best_fit_planck}. This signifies the fact that the OTM has a close resemblance to the usual LCDM model. However, we observe that the amount of cold dark matter $\Omega_ch^2$ for the oscillatory tracker model is comparatively lower than the base LCDM model. This slight decrease in the amount of matter content of the universe has a direct effect on the CMB sector and matter sector, which we further investigate in section~\ref{Sec:effects}. It is also worth to note that the hubble parameter value $H_0$ for oscillatory tracker model is consistent with LCDM Planck 2018 results ($H_0 = (67.27 \pm 0.6)$\,Km s$^{-1}$Mpc$^{-1}$) within $1\,\sigma$.  Moreover, the flat posteriors for the model parameters $\alpha$ and $c$ indicate that, there is a broad range of parameter values for $\alpha$ and $c$ for which the OTM is consistent with the set of observational data presented in section~\ref{Sec:cmbdata}. 
\begin{figure}
	\centering
	\includegraphics[width=0.49\textwidth]{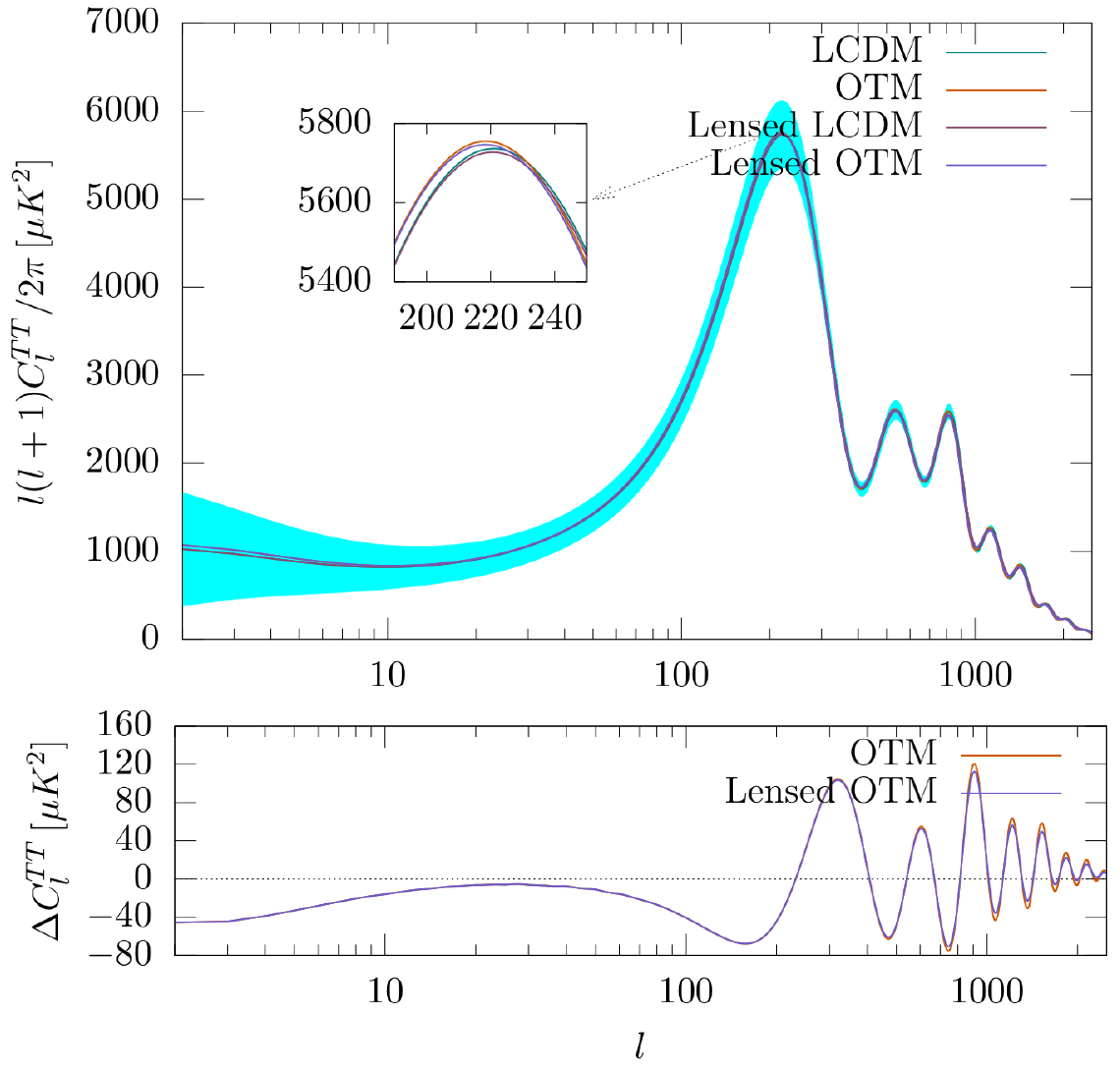}
	\caption{\small Top panel: Figure showing the comparison of CMB temperature (TT) power spectrum of OTM with respect to the base LCDM model for both lensed and unlensed cases. The inset shows the zoomed-in versions of the peak and the cyan band represents the cosmic variance corresponding to $C_l^{TT}$ of LCDM. Due to the presence of the scalar field, the peak positions of OTM power spectrum are marginally higher than the base LCDM. Bottom Panel: Figure showing the deviations of the CMB TT power spectrum of the OTM from the base LCDM for both lensed and unlensed cases. \normalsize}
	\label{fig:cmb_TT}
\end{figure}

%-------------------------
%-------------------------
\section{Effects on Observable Probes} \label{Sec:effects}
%-------------------------
%-------------------------

In this section, in order to study the impact of oscillatory tracker dark energy model on different observable quantities, we focus on its effects on CMB power spectra (TT, EE, TE), matter power spectrum and $f \sigma_8$. For this analysis, we make use of the modified version of the public available boltzmann code CLASS~\citep{Blas_2011,lesgourgues2011cosmic} with the best-fit values of the cosmological parameters from table~\ref{tab:best_fit}. On the other hand, for the case of the LCDM model, we implement the cosmological parameter values from Planck 2018 data~\citep{2020} (see table~\ref{tab:best_fit_planck}) in CLASS and hereon, the LCDM model is our base model.  
\subsection{CMB Sector}\label{Sec:cmbeffects}
The CMB radiation has a significant role in understanding our universe as its an open window to the early universe. This radiation field is nearly isotropic and exhibits almost a perfect black body spectrum at a temperature of $2.73K$~\citep{2020}. The numerical evaluation of CMB temperature and polarization power spectra is implemented in the CLASS code. The comparison of CMB TT power spectrum of oscillatory tracker model with respect to base LCDM is shown with and without lensing effects in the top panel of figure~\ref{fig:cmb_TT}. Here the inset shows the zoomed-in versions of the first peak of the CMB TT power spectrum. From our analysis of oscillatory tracker model with the observational datasets given in section~\ref{Sec:cmbdata}, we find that the cold dark matter density $\Omega_c h^2$ for OTM is slightly lower than the base LCDM model (see table~\ref{tab:best_fit} and table~\ref{tab:best_fit_planck}).

\begin{figure}
	\centering
	\includegraphics[width=0.49\textwidth]{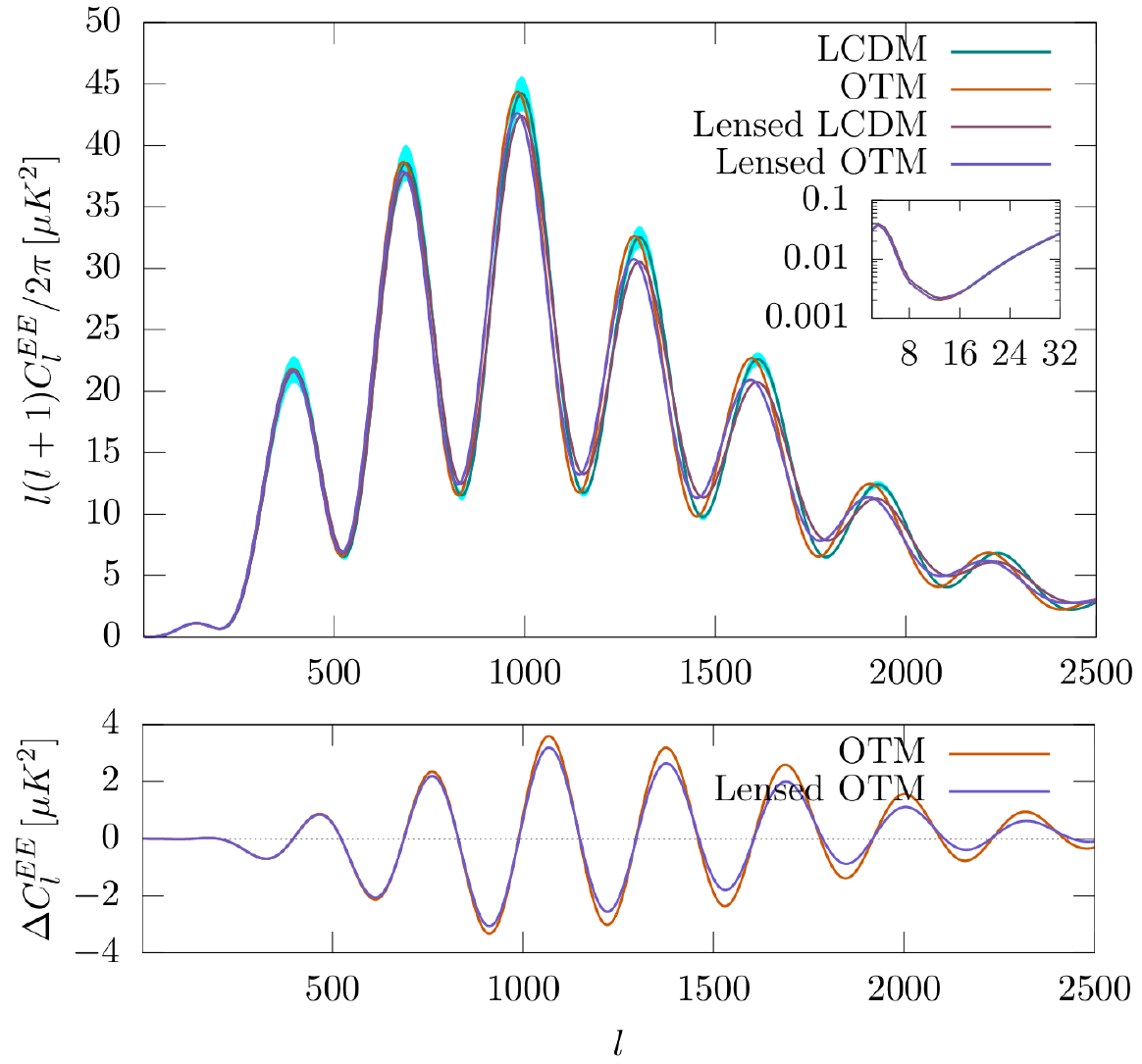}
	\caption{\small Top panel: Figure showing the comparison of CMB polarization (EE) power spectrum of OTM with respect to the base LCDM model for both lensed and unlensed cases. The inset shows the zoomed-in versions of the low-$\ell$ modes, and the cyan band represents the cosmic variance corresponding to $C_l^{EE}$ of LCDM. The peak positions of both OTM and LCDM are slightly lowered when the lensing effects are taken into account, and its effects are noticeably higher at higher multipoles. Bottom panel: Figure showing the deviations of  CMB polarization (EE) power spectrum from the base LCDM for both lensed and unlensed cases. \normalsize
		\label{fig:cmb_EE}
	}
\end{figure}

The amount of dark matter plays a significant role in determining the time at which the universe transitioned from radiation-dominated epoch to matter-dominated epoch. So a lower dark matter density in the oscillatory tracker model results in delaying this transition epoch. As a result, the universe enters matter domination later compared to the usual LCDM model. As the universe becomes more radiation dominated in the early phases of the evolution of OTM, it affects the gravitational potential wells. It thus causes an increase in the so-called 'radiation driving effect'~\citep{Hu_2002} which manifests as a rise in the acoustic peaks of the CMB power spectrum. Due to this driving effect, the peak positions of the CMB TT power spectrum for the oscillatory tracker model is marginally higher than the base LCDM model (see bottom panel of the figure~\ref{fig:cmb_TT}). The presence of the scalar field also increases the low-$\ell$ modes of the CMB TT power spectrum through the Integrated Sachs-Wolfe (ISW) effect. Moreover, with matter-radiation equality occurring later in OTM, the decay of the gravitational potentials goes further beyond the decoupling. As a result, the stronger ISW effects also contribute coherently to the height of the first peak (see inset of figure~\ref{fig:cmb_TT}). As the cosmic variance is larger in the lower multipole regions, the additional effects introduced by the scalar field is difficult to distinguish from the predictions of the LCDM model. It will be an interesting future work to observationally detect such minor changes for the case of OTM from the LCDM predictions in the cosmic variance dominated low multipole regions. The bottom panel of figure~\ref{fig:cmb_TT} shows the deviations of the CMB TT power spectrum of the oscillatory tracker model from the base LCDM model for both lensed and unlensed cases. It is evident that the deviations of OTM from the LCDM are dominant at the acoustic peaks, and it decreases for higher multipoles ranges.

\begin{figure}
	\centering
	\includegraphics[width=0.49\textwidth]{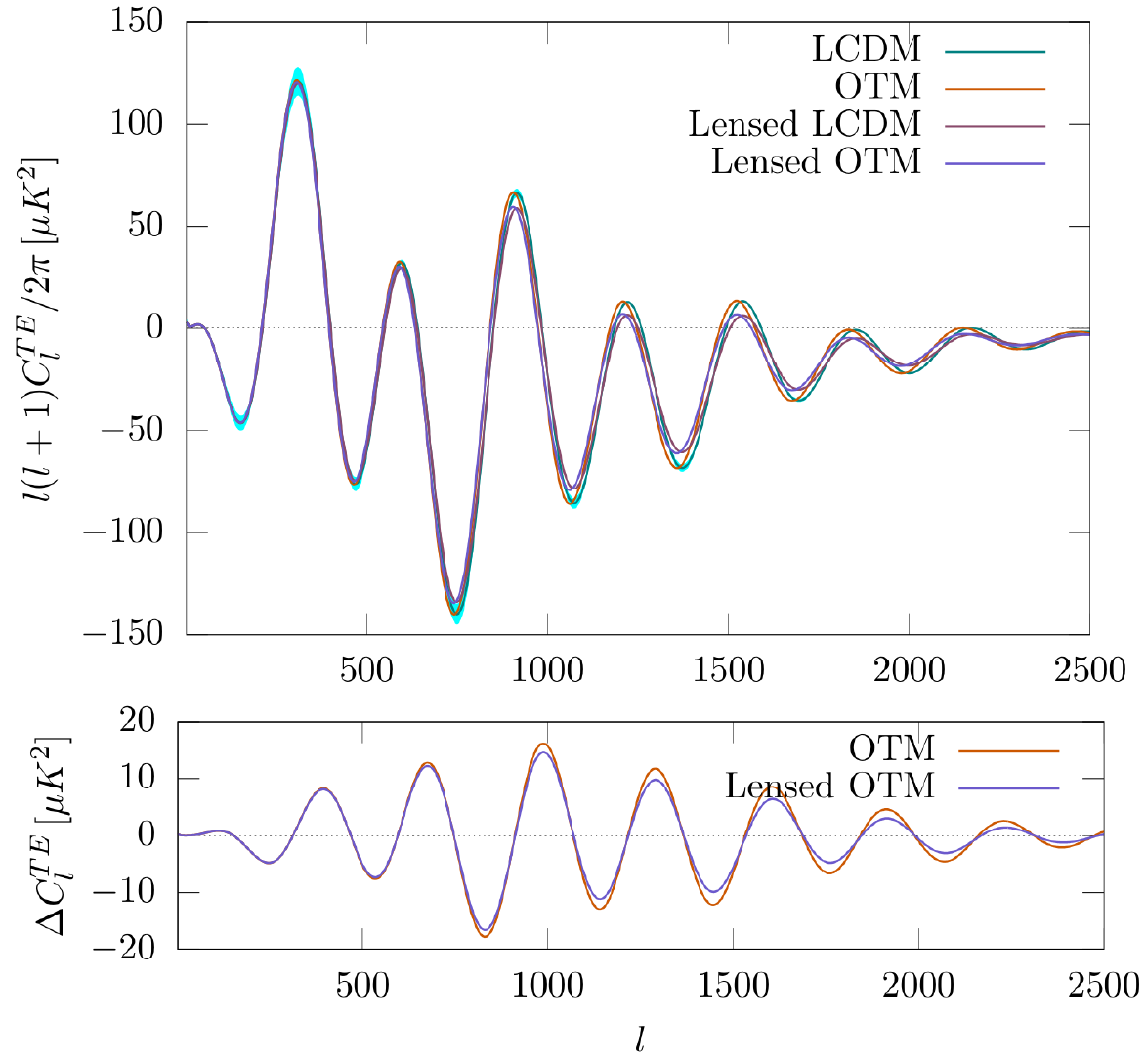}
	\caption{\small Top panel: Figure showing the comparison of CMB temperature-polarization (TE) power spectrum of OTM with respect to the base LCDM model for both lensed and unlensed cases. The cyan band represents the cosmic variance corresponding to $C_l^{TE}$ of LCDM. The lensing effects slightly reduce the peak positions of the CMB TE power spectrum, and the effects are noticeably higher at higher multipoles. Bottom panel: Figure showing the CMB TE power spectrum deviations from the base LCDM for lensed and unlensed cases.\normalsize 
		\label{fig:cmb_TE}
	}
\end{figure}

The comparison of CMB EE and TE power spectra of oscillatory tracker dark energy model with respect to the base LCDM model is shown with and without lensing effects in figure~\ref{fig:cmb_EE} and figure~\ref{fig:cmb_TE} respectively. The polarization peaks are at the troughs of the CMB temperature power spectrum. The effects of scalar field on the low-$\ell$ modes of the CMB EE power spectrum is shown in the inset in the figure~\ref{fig:cmb_EE}. As the low-$\ell$ modes of EE power spectrum of OTM closely match LCDM, the reionization history in both models are not altered to a great extent. This is due to the fact that the optical depth to the reionization $\tau_{reio}$ for OTM (see table~\ref{tab:best_fit}) is consistent with LCDM Planck 2018~\citep{2020} within $1\, \sigma$. The peak heights of CMB temperature and polarization power spectra are marginally lowered when the lensing effects are considered.  Moreover, it is also visible that the effect of lensing is noticeably higher at higher multipoles.

\begin{figure}
	\centering
	\includegraphics[width=0.49\textwidth]{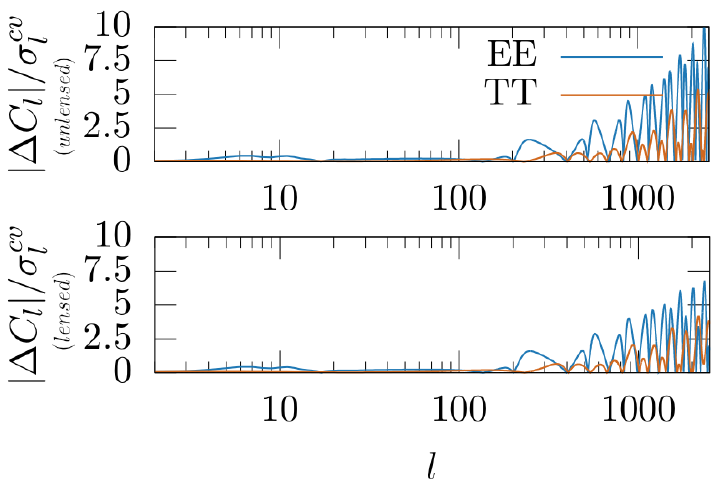}
	\caption{\small Top panel: Figure showing the fraction of absolute difference of CMB angular power spectra (TT and EE)  between the OTM and LCDM models with the LCDM predicted cosmic variance errors without CMB weak lensing taken into considerations. Bottom panel: Same as top panel, but power spectra are obtained taking into account CMB lensing effects as well. Rapidly increasing deviations between the OTM and LCDM models are predicted at larger multipoles at the locations of peaks and troughs of the acoustic oscillations of the two spectra. Comparing both TT and EE, it is interesting to note that the fractional change for the case of EE  is somewhat larger than the TT case at large multipoles. \normalsize
		\label{fig:cv}
	}
\end{figure}

\subsection{Matter Sector}\label{Sec:cmbeffects}

\begin{figure}
	\centering
	\includegraphics[width=0.49\textwidth]{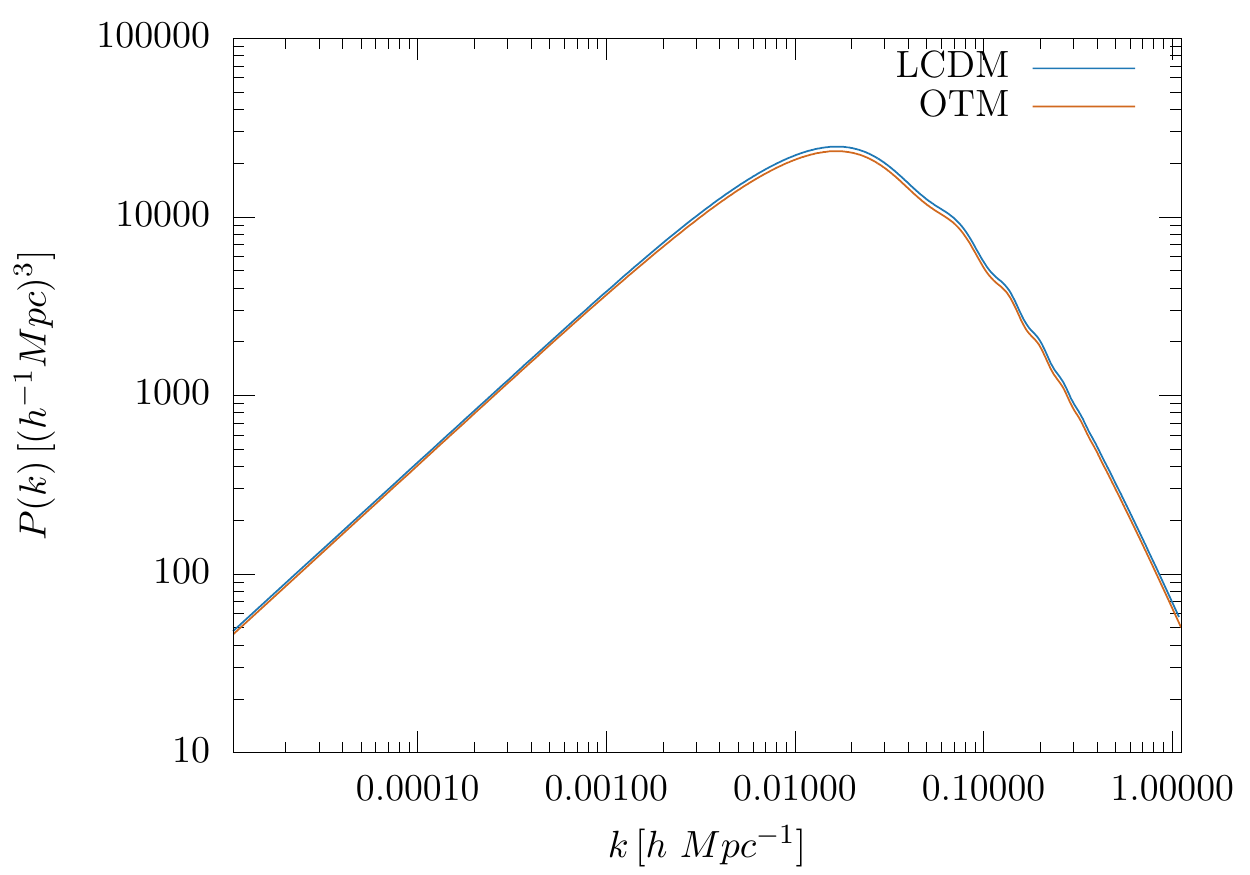}
	\caption{\small
		The figure showing the variation of matter power spectrum $P(k)$ as a function of comoving wave number $k$ for both OTM and base LCDM model. The presence of scalar field sightly reduces the matter content of the universe; as a result, the matter power spectrum of OTM is slightly lower than the base LCDM model. \normalsize
		\label{fig:matter_ps}
	}
\end{figure}

The observed cosmic structures are the results of the amplification of primordial density fluctuations by gravitational instability. The power spectrum of matter density fluctuations plays a significant role in understanding the dynamics of our universe. The matter power spectrum~\citep{dodelson} can be written as,
\begin{equation} \label{power}
	P\left(k,a\right)= A_s \,k^{n_s} T^2\left(k\right) D^2\left(a\right),
\end{equation}
where $A_s$ is the scalar primordial power spectrum amplitude, $T(k)$ is the matter transfer function, $n_s$ is the spectral index and $D(a)=\frac{\delta_m(a)}{\delta_m(a=1)}$ is the normalized density contrast. The combined effects of the complementary actions of the outward push by the radiation pressure and the inward pull by the gravity, which are responsible for the acoustic oscillations in the CMB, also determine the power spectrum of non-relativistic matter. From our analysis of the oscillatory tracker model with observational datasets given in section~\ref{Sec:cmbdata}, we find that because of the presence of scalar field; there is a minor decrease in the amount of cold dark matter $\Omega_{c}h^2$ present in the universe compared to the base LCDM model (see table~\ref{tab:best_fit} and table~\ref{tab:best_fit_planck}). A decrease in the matter content of the universe in turn, reduces the matter power spectrum compared to the base LCDM model. This effect of the scalar field that decreases matter power spectrum is shown in figure~\ref{fig:matter_ps}. Any dynamical effect that reduces the amplitude of the matter power spectrum corresponds to a decay in the Newtonian potential that boosts the level of anisotropy~\citep{Hu_2002}. Thus a decrease in the matter content of the universe due to the presence of scalar field drives the matter power spectrum down and the CMB spectrum up.

\begin{figure}
	\centering
	\includegraphics[width=0.49\textwidth]{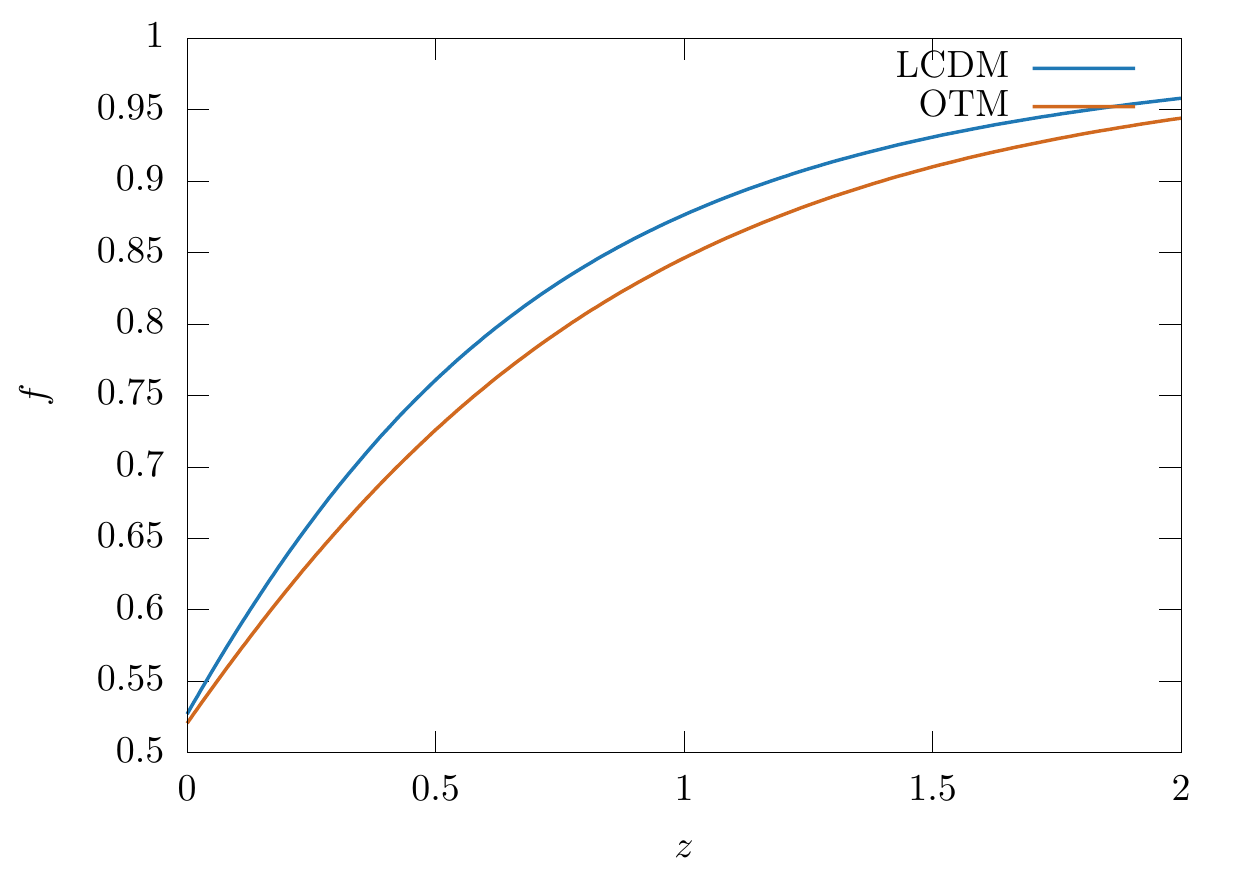}
	\caption{\small Figure showing the variation of linear growth rate $f$ as a function of redshift $z$ in the low redshift regime. The linear growth rate for OTM is slightly lower compared to the base LCDM model. \normalsize
		\label{fig:growth_rate}
	}
\end{figure}

Another great tool to differentiate various dark energy models based on the growth of large scale structures is the linear growth rate. One can define the linear growth rate as,
\begin{equation} \label{eq:growth_rate}
	f(a)= \frac{d \ln \delta_{m}}{d \ln a}~=~ \frac{a}{\delta_{m}(a)}\frac{d \delta_{m}}{d \,a}~.
\end{equation}

\begin{figure}
	\centering
	\includegraphics[width=0.5\textwidth]{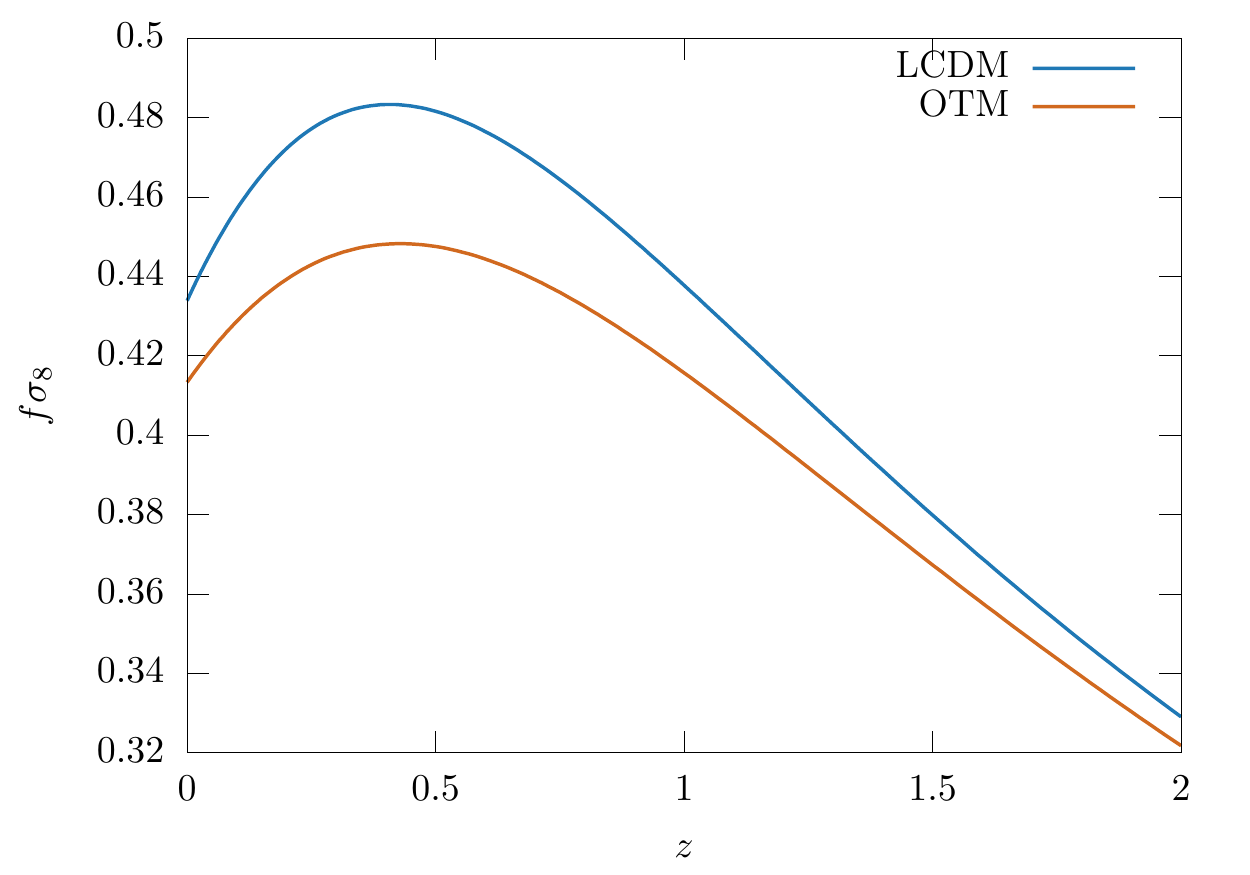}
	\caption{\small Figure showing the variation of $f\sigma_8$ as a function of redshift $z$ in the low redshift regime. At the present epoch $z=0$, the $f \sigma_8$ of OTM is lower than the base LCDM model. \normalsize
		\label{fig:fs8}
	}	
\end{figure}

A more powerful and dependable observational quantity that is measured by redshift surveys is the product of $f(a)\sigma_8$~\citep{Percival_2009}, where $\sigma_8$ is the root-mean-square (rms) fluctuations of the linear density field within the sphere of radius $R = 8 \,h^{-1}$Mpc. In the linear regime, $\sigma_8$ and $f \sigma_8$ can be read as~\citep{PhysRevD.77.023504, Song_2009,Huterer_2015,Ishak_2018},
\begin{eqnarray}
	\sigma_8(z) = \sigma_8(z=0)\frac{\delta_{m}(z)}{\delta_{m}(z=0)},
\end{eqnarray}
and
\begin{equation} \label{eq:fsigma}
	f \sigma_8(z) \equiv f(z)\sigma_8(z) =-(1+z)\frac{\sigma_8(z=0)}{\delta_{m}(z=0)}\frac{d \delta_{m}}{d \,z}~,
\end{equation}
where $\sigma_8(z=0)$ is the value of the rms fluctuations of the linear density field at $z=0$. The redshift $z$ is related to the scale factor $a$ as $z=\frac{a_0}{a}\,-1$ where $a_0$ is the present value of the scale factor. We obtained the $\sigma_8$ value (see table~\ref{tab:sigma8}) of OTM using the best-fit parameters given in table~\ref{tab:best_fit} whereas for LCDM model we used the mean values given in table~\ref{tab:best_fit_planck}. As $f\sigma_8$ is a more reliable quantity~\citep{Percival_2009}, it gives a better insight into the growth of the density perturbations. At low redshift, both the linear growth rate $f$ and $f\sigma_8$ are independent of the wave number $k$. So we consider the redshift in the ranges $z=0$ to $z=2$. The variation of the linear growth rate and $f\sigma_8$ as a function of redshift $z$ is shown in figure~\ref{fig:growth_rate} and figure~\ref{fig:fs8} respectively. The linear growth rate for the case of the oscillatory dark energy model is slightly lower than the base LCDM model. Both linear growth rate $f$ and $f\sigma_8$ are lower for the oscillatory model at redshift $z=0$. The difference in the matter power spectrum is manifested in the amplitude of $\sigma_8$ given in table~\ref{tab:sigma8} and hence in $f\sigma_8$ shown in figure~\ref{fig:fs8}. Moreover, the lower value of $\sigma_8$ in OTM also results in the reduction of the clustering of galaxies compared to base LCDM.
\begin{table}
	\centering
	\begin{tabular}{lc} 
		\hline 
		Model   &   $\sigma_8$ \\ 
		\hline 
		\hline
		OTM
		&  $0.7939$
		\\
		LCDM
		& 	$0.8231$
		\\
		\hline 
	\end{tabular} 
	\caption{\small
		The values of $\sigma_8$ at redshift $z=0$ for the OTM and the LCDM model. \normalsize
		\label{tab:sigma8}
	}
\end{table}

%%%%-------

%%

\section{Conclusion} \label{Sec:Conclusions}

The dark energy models with tracker properties have gained great attention over the years because of their ability to alleviate the cosmic coincidence problem. In this scenario, the scalar field which drives the accelerated expansion of the universe can reach the present value of dark energy density from a wide range of initial conditions. In this work, we investigated and extended the study of the oscillatory tracker dark energy model~\citep{Cede_o_2019}, which belongs to a family of tracking dark energy models known as $\alpha$-attractors  against various observational data consisting of CMB, BAO and type 1a supernovae data. The oscillatory tracker dark energy model, which is very much favoured over other $\alpha$-attractor dark energy models~\citep{Cede_o_2019} has a large initial attractor basin. As a result, the present observed dark energy density can be obtained from a very large range of initial conditions, which provides a solution to the so-called 'fine-tuning' problem.
%During our analysis, since we have obtained same results for all the scalar field initial conditions %$(\phi_i,\dot{\phi_i}) = (0\,to\,10, 0)$, we have fixed the initial conditions as %$(\phi_i,\dot{\phi_i}) = (10, 0)$. At earlier times the field gets freezes because of Hubble %friction, so we fixed the scalar field velocity $\dot{\phi_i}$ to zero. 

By comparing the OTM against various observational data (CMB, BAO, type 1a supernovae), we constrained the parameters of the model in section~\ref{Sec:Methods}. From the quantitative results given in table~\ref{tab:best_fit}, it is interesting to note that the standard cosmological parameters are in good agreement with the LCDM Planck 2018 results~\citep{2020}. This signifies the fact that the OTM has a close resemblance to the usual LCDM model. Moreover, the flat posteriors of the parameters of oscillatory tracker model $\alpha$ and $c$ indicate that there is a broad range of allowed values for these parameters for which the OTM is consistent with the set of observational data presented in section~\ref{Sec:cmbdata}. 

After obtaining the best-fit model parameters, in section~\ref{Sec:effects} we investigated the effect of the OTM on CMB temperature and polarization power spectra, matter power spectrum and the $f\sigma_8$. Though the oscillatory tracker model and usual LCDM are qualitatively very similar, they are not really overlapping. The less dark matter content in the oscillatory tracker model leads to a later epoch of matter-radiation equality and more decay of gravitational potentials. This causes an increase in the radiation driving effects and results in the rise of the amplitudes of the acoustic peaks in the oscillatory tracker model. The fraction of absolute difference of CMB angular power spectra (TT and EE) between OTM and LCDM models with the LCDM predicted cosmic variance errors also suggest an increase in the deviations at larger multipoles at the locations of peaks and troughs of the acoustic oscillations of the two spectra.
The effect of the decrease in matter content of the universe due to the presence of the scalar field is also seen as a decrease in the power of matter power spectrum compared to the base LCDM model. 
The comparison between the CMB and the large scale structure is important, since it breaks the degeneracies between effects due to deviations from power law initial conditions and the dynamics of the matter and energy content of the universe~\citep{Hu_2002}. Any dynamical effect which reduces the matter power spectrum corresponds to the decay of gravitational potential wells that boosts the amplitude of the acoustic peaks in the CMB power spectrum. As a result, the decrease in matter content of the universe in OTM drives the matter power spectrum down and the CMB power spectrum up. Moreover, our investigation on the linear growth rate $f$ and $f \sigma_8$ at a low redshift regime indicates that the $f$ and $f\sigma_8$ are slightly lower than the base LCDM model. The lower value of $\sigma_8$ of the oscillatory tracker dark energy model signifies that the presence of scalar field reduces the galaxy clustering compared to the base LCDM model. 

Our analysis on the CMB sector and matter sector suggest that the oscillatory tracker dark energy model can be equally viable to the usual LCDM model given the current set of observational data used in this work. It will be interesting to investigate the future generation more sensitive observations like HETDEX~\citep{hill2008hobbyeberly}, WFIRST~\citep{spergel2015widefield} and LSST~\citep{Ivezi__2019} in detail so as to detect the small differences between the OTM and LCDM models reported in this article in order to constrain the models further.

\section*{Acknowledgements}
	Computations were carried out on Kanad, the high performance computation facility of IISER Bhopal, India. AJ acknowledges financial support from the Ministry of Human Resource and Development, Government of India via Institute fellowship at IISER Bhopal.

\section{Data Availability}
The data underlying this article will be shared on reasonable request to the corresponding author.



\bibliographystyle{mnras}
\bibliography{ms} % if your bibtex file is called example.bib

\begin{thebibliography}{}
\makeatletter
\relax
\def\mn@urlcharsother{\let\do\@makeother \do\$\do\&\do\#\do\^\do\_\do\%\do\~}
\def\mn@doi{\begingroup\mn@urlcharsother \@ifnextchar [ {\mn@doi@}
  {\mn@doi@[]}}
\def\mn@doi@[#1]#2{\def\@tempa{#1}\ifx\@tempa\@empty \href
  {http://dx.doi.org/#2} {doi:#2}\else \href {http://dx.doi.org/#2} {#1}\fi
  \endgroup}
\def\mn@eprint#1#2{\mn@eprint@#1:#2::\@nil}
\def\mn@eprint@arXiv#1{\href {http://arxiv.org/abs/#1} {{\tt arXiv:#1}}}
\def\mn@eprint@dblp#1{\href {http://dblp.uni-trier.de/rec/bibtex/#1.xml}
  {dblp:#1}}
\def\mn@eprint@#1:#2:#3:#4\@nil{\def\@tempa {#1}\def\@tempb {#2}\def\@tempc
  {#3}\ifx \@tempc \@empty \let \@tempc \@tempb \let \@tempb \@tempa \fi \ifx
  \@tempb \@empty \def\@tempb {arXiv}\fi \@ifundefined
  {mn@eprint@\@tempb}{\@tempb:\@tempc}{\expandafter \expandafter \csname
  mn@eprint@\@tempb\endcsname \expandafter{\@tempc}}}

\bibitem[\protect\citeauthoryear{Abazajian et~al.,}{Abazajian
  et~al.}{2015}]{Abazajian_2015}
Abazajian K.,  et~al., 2015, \mn@doi [Astroparticle Physics]
  {10.1016/j.astropartphys.2014.05.013}, 63, 55–65

\bibitem[\protect\citeauthoryear{Abramo \& Finelli}{Abramo \&
  Finelli}{2003}]{Abramo_2003}
Abramo L.,  Finelli F.,  2003, \mn@doi [Physics Letters B]
  {10.1016/j.physletb.2003.09.065}, 575, 165–171

\bibitem[\protect\citeauthoryear{Aghanim et~al.,}{Aghanim et~al.}{2020}]{2020}
Aghanim N.,  et~al., 2020, \mn@doi [Astronomy $\&$ Astrophysics]
  {10.1051/0004-6361/201833910}, 641, A6

\bibitem[\protect\citeauthoryear{Aguirregabiria \& Lazkoz}{Aguirregabiria \&
  Lazkoz}{2004}]{Aguirregabiria_2004}
Aguirregabiria J.,  Lazkoz R.,  2004, \mn@doi [Physical Review D]
  {10.1103/physrevd.69.123502}, 69

\bibitem[\protect\citeauthoryear{Akrami, Kallosh, Linde  \& Vardanyan}{Akrami
  et~al.}{2018}]{Akrami_2018}
Akrami Y.,  Kallosh R.,  Linde A.,   Vardanyan V.,  2018, \mn@doi [Journal of
  Cosmology and Astroparticle Physics] {10.1088/1475-7516/2018/06/041}, 2018,
  041–041

\bibitem[\protect\citeauthoryear{{Alam} et~al.}{{Alam}
  et~al.}{2017}]{2017MNRAS.470.2617A}
{Alam} S.,  et~al., 2017, \mn@doi [\mnras] {10.1093/mnras/stx721}, \href
  {https://ui.adsabs.harvard.edu/abs/2017MNRAS.470.2617A} {470, 2617}

\bibitem[\protect\citeauthoryear{Armendariz-Picon, Mukhanov  \&
  Steinhardt}{Armendariz-Picon et~al.}{2000}]{Armendariz_Picon_2000}
Armendariz-Picon C.,  Mukhanov V.,   Steinhardt P.~J.,  2000, \mn@doi [Physical
  Review Letters] {10.1103/physrevlett.85.4438}, 85, 4438–4441

\bibitem[\protect\citeauthoryear{Armendariz-Picon, Mukhanov  \&
  Steinhardt}{Armendariz-Picon et~al.}{2001}]{Armendariz_Picon_2001}
Armendariz-Picon C.,  Mukhanov V.,   Steinhardt P.~J.,  2001, \mn@doi [Physical
  Review D] {10.1103/physrevd.63.103510}, 63

\bibitem[\protect\citeauthoryear{Bag, Mishra  \& Sahni}{Bag
  et~al.}{2018}]{Bag_2018}
Bag S.,  Mishra S.~S.,   Sahni V.,  2018, \mn@doi [Journal of Cosmology and
  Astroparticle Physics] {10.1088/1475-7516/2018/08/009}, 2018, 009–009

\bibitem[\protect\citeauthoryear{Bagla, Jassal  \& Padmanabhan}{Bagla
  et~al.}{2003}]{Bagla_2003}
Bagla J.~S.,  Jassal H.~K.,   Padmanabhan T.,  2003, \mn@doi [Physical Review
  D] {10.1103/physrevd.67.063504}, 67

\bibitem[\protect\citeauthoryear{Barreiro, Copeland  \& Nunes}{Barreiro
  et~al.}{2000}]{Barreiro:1999zs}
Barreiro T.,  Copeland E.~J.,   Nunes N.~J.,  2000, \mn@doi [Phys. Rev. D]
  {10.1103/PhysRevD.61.127301}, 61, 127301

\bibitem[\protect\citeauthoryear{{Beutler} et~al.,}{{Beutler}
  et~al.}{2011}]{2011MNRAS.416.3017B}
{Beutler} F.,  et~al., 2011, \mn@doi [\mnras]
  {10.1111/j.1365-2966.2011.19250.x}, \href
  {https://ui.adsabs.harvard.edu/abs/2011MNRAS.416.3017B} {416, 3017}

\bibitem[\protect\citeauthoryear{Beutler, Blake, Koda, Marín, Seo, Cuesta  \&
  Schneider}{Beutler et~al.}{2015}]{Beutler_2015}
Beutler F.,  Blake C.,  Koda J.,  Marín F.~A.,  Seo H.-J.,  Cuesta A.~J.,
  Schneider D.~P.,  2015, \mn@doi [Monthly Notices of the Royal Astronomical
  Society] {10.1093/mnras/stv1943}, 455, 3230–3248

\bibitem[\protect\citeauthoryear{Bezrukov \& Shaposhnikov}{Bezrukov \&
  Shaposhnikov}{2008}]{Bezrukov_2008}
Bezrukov F.,  Shaposhnikov M.,  2008, \mn@doi [Physics Letters B]
  {10.1016/j.physletb.2007.11.072}, 659, 703–706

\bibitem[\protect\citeauthoryear{Blake et~al.,}{Blake
  et~al.}{2012}]{Blake_2012}
Blake C.,  et~al., 2012, \mn@doi [Monthly Notices of the Royal Astronomical
  Society] {10.1111/j.1365-2966.2012.21473.x}, 425, 405–414

\bibitem[\protect\citeauthoryear{Blas, Lesgourgues  \& Tram}{Blas
  et~al.}{2011}]{Blas_2011}
Blas D.,  Lesgourgues J.,   Tram T.,  2011, \mn@doi [Journal of Cosmology and
  Astroparticle Physics] {10.1088/1475-7516/2011/07/034}, 2011, 034–034

\bibitem[\protect\citeauthoryear{Brax \& Davis}{Brax \&
  Davis}{2015}]{Brax_2015}
Brax P.,  Davis A.-C.,  2015, \mn@doi [Journal of Cosmology and Astroparticle
  Physics] {10.1088/1475-7516/2015/10/042}, 2015, 042–042

\bibitem[\protect\citeauthoryear{Brown, Cottrell, Shiu  \& Soler}{Brown
  et~al.}{2016}]{Brown_2016}
Brown J.,  Cottrell W.,  Shiu G.,   Soler P.,  2016, \mn@doi [Journal of High
  Energy Physics] {10.1007/jhep10(2016)025}, 2016

\bibitem[\protect\citeauthoryear{Caldwell}{Caldwell}{2002}]{Caldwell_2002}
Caldwell R.,  2002, \mn@doi [Physics Letters B]
  {10.1016/s0370-2693(02)02589-3}, 545, 23–29

\bibitem[\protect\citeauthoryear{Caldwell, Kamionkowski  \& Weinberg}{Caldwell
  et~al.}{2003}]{Caldwell_2003}
Caldwell R.~R.,  Kamionkowski M.,   Weinberg N.~N.,  2003, \mn@doi [Physical
  Review Letters] {10.1103/physrevlett.91.071301}, 91

\bibitem[\protect\citeauthoryear{Cedeño, Montiel, Hidalgo  \& German}{Cedeño
  et~al.}{2019}]{Cede_o_2019}
Cedeño F. X.~L.,  Montiel A.,  Hidalgo J.~C.,   German G.,  2019, \mn@doi
  [Journal of Cosmology and Astroparticle Physics]
  {10.1088/1475-7516/2019/08/002}, 2019, 002–002

\bibitem[\protect\citeauthoryear{Cervantes-Cota \& Dehnen}{Cervantes-Cota \&
  Dehnen}{1995}]{Cervantes_Cota_1995}
Cervantes-Cota J.,  Dehnen H.,  1995, \mn@doi [Nuclear Physics B]
  {10.1016/0550-3213(95)00128-x}, 442, 391–409

\bibitem[\protect\citeauthoryear{Chiba, Okabe  \& Yamaguchi}{Chiba
  et~al.}{2000}]{Chiba_2000}
Chiba T.,  Okabe T.,   Yamaguchi M.,  2000, \mn@doi [Physical Review D]
  {10.1103/physrevd.62.023511}, 62

\bibitem[\protect\citeauthoryear{Chiba, De~Felice  \& Tsujikawa}{Chiba
  et~al.}{2013}]{Chiba_2013}
Chiba T.,  De~Felice A.,   Tsujikawa S.,  2013, \mn@doi [Physical Review D]
  {10.1103/physrevd.87.083505}, 87

\bibitem[\protect\citeauthoryear{Clifton, Ferreira, Padilla  \&
  Skordis}{Clifton et~al.}{2012}]{Clifton_2012}
Clifton T.,  Ferreira P.~G.,  Padilla A.,   Skordis C.,  2012, \mn@doi [Physics
  Reports] {10.1016/j.physrep.2012.01.001}, 513, 1–189

\bibitem[\protect\citeauthoryear{Conlon}{Conlon}{2012}]{Conlon_2012}
Conlon J.~P.,  2012, \mn@doi [Journal of Cosmology and Astroparticle Physics]
  {10.1088/1475-7516/2012/01/033}, 2012, 033–033

\bibitem[\protect\citeauthoryear{Copeland, Garousi, Sami  \&
  Tsujikawa}{Copeland et~al.}{2005}]{Copeland_2005}
Copeland E.~J.,  Garousi M.~R.,  Sami M.,   Tsujikawa S.,  2005, \mn@doi
  [Physical Review D] {10.1103/physrevd.71.043003}, 71

\bibitem[\protect\citeauthoryear{Damour, Piazza  \& Veneziano}{Damour
  et~al.}{2002}]{Damour_2002}
Damour T.,  Piazza F.,   Veneziano G.,  2002, \mn@doi [Physical Review Letters]
  {10.1103/physrevlett.89.081601}, 89

\bibitem[\protect\citeauthoryear{Dimitrijevic, Dragovich, Grujic  \&
  Rakic}{Dimitrijevic et~al.}{2013}]{Dimitrijevic_2013}
Dimitrijevic I.,  Dragovich B.,  Grujic J.,   Rakic Z.,  2013, \mn@doi [Lie
  Theory and Its Applications in Physics] {10.1007/978-4-431-54270-4_17}, p.
  251–259

\bibitem[\protect\citeauthoryear{Dodelson}{Dodelson}{2003}]{dodelson}
Dodelson S.,  2003, Modern Cosmology

\bibitem[\protect\citeauthoryear{Durrive, Ooba, Ichiki  \& Sugiyama}{Durrive
  et~al.}{2018}]{Durrive_2018}
Durrive J.-B.,  Ooba J.,  Ichiki K.,   Sugiyama N.,  2018, \mn@doi [Physical
  Review D] {10.1103/physrevd.97.043503}, 97

\bibitem[\protect\citeauthoryear{Eisenstein et~al.,}{Eisenstein
  et~al.}{2005}]{Eisenstein_2005}
Eisenstein D.~J.,  et~al., 2005, \mn@doi [The Astrophysical Journal]
  {10.1086/466512}, 633, 560–574

\bibitem[\protect\citeauthoryear{Ferrara, Kallosh, Linde, Marrani  \&
  Van~Proeyen}{Ferrara et~al.}{2010}]{PhysRevD.82.045003}
Ferrara S.,  Kallosh R.,  Linde A.,  Marrani A.,   Van~Proeyen A.,  2010,
  \mn@doi [Phys. Rev. D] {10.1103/PhysRevD.82.045003}, 82, 045003

\bibitem[\protect\citeauthoryear{Ferreira \& Joyce}{Ferreira \&
  Joyce}{1997}]{Ferreira:1997au}
Ferreira P.~G.,  Joyce M.,  1997, \mn@doi [Phys. Rev. Lett.]
  {10.1103/PhysRevLett.79.4740}, 79, 4740

\bibitem[\protect\citeauthoryear{Ferreira \& Joyce}{Ferreira \&
  Joyce}{1998}]{Ferreira:1997hj}
Ferreira P.~G.,  Joyce M.,  1998, \mn@doi [Phys. Rev. D]
  {10.1103/PhysRevD.58.023503}, 58, 023503

\bibitem[\protect\citeauthoryear{Flauger, McAllister, Pajer, Westphal  \&
  Xu}{Flauger et~al.}{2010}]{Flauger_2010}
Flauger R.,  McAllister L.,  Pajer E.,  Westphal A.,   Xu G.,  2010, \mn@doi
  [Journal of Cosmology and Astroparticle Physics]
  {10.1088/1475-7516/2010/06/009}, 2010, 009–009

\bibitem[\protect\citeauthoryear{García-García, Linder, Ruíz-Lapuente  \&
  Zumalacárregui}{García-García et~al.}{2018}]{Garc_a_Garc_a_2018}
García-García C.,  Linder E.~V.,  Ruíz-Lapuente P.,   Zumalacárregui M.,
  2018, \mn@doi [Journal of Cosmology and Astroparticle Physics]
  {10.1088/1475-7516/2018/08/022}, 2018, 022–022

\bibitem[\protect\citeauthoryear{Gelman \& Rubin}{Gelman \&
  Rubin}{1992}]{Gelman:1992zz}
Gelman A.,  Rubin D.~B.,  1992, \mn@doi [Statist. Sci.]
  {10.1214/ss/1177011136}, 7, 457

\bibitem[\protect\citeauthoryear{Goncharov \& Linde}{Goncharov \&
  Linde}{1984a}]{Goncharov:1984jlb}
Goncharov A.~S.,  Linde A.~D.,  1984a, Sov. Phys. JETP, 59, 930

\bibitem[\protect\citeauthoryear{Goncharov \& Linde}{Goncharov \&
  Linde}{1984b}]{Goncharov:1983mw}
Goncharov A.~B.,  Linde A.~D.,  1984b, \mn@doi [Phys. Lett. B]
  {10.1016/0370-2693(84)90027-3}, 139, 27

\bibitem[\protect\citeauthoryear{Guo \& Zhang}{Guo \& Zhang}{2004}]{Guo_2004}
Guo Z.-K.,  Zhang Y.-Z.,  2004, \mn@doi [Journal of Cosmology and Astroparticle
  Physics] {10.1088/1475-7516/2004/08/010}, 2004, 010–010

\bibitem[\protect\citeauthoryear{Guth}{Guth}{1981}]{PhysRevD.23.347}
Guth A.~H.,  1981, \mn@doi [Phys. Rev. D] {10.1103/PhysRevD.23.347}, 23, 347

\bibitem[\protect\citeauthoryear{Hill et~al.,}{Hill
  et~al.}{2008}]{hill2008hobbyeberly}
Hill G.~J.,  et~al., 2008, The Hobby-Eberly Telescope Dark Energy Experiment
  (HETDEX): Description and Early Pilot Survey Results (\mn@eprint {arXiv}
  {0806.0183})

\bibitem[\protect\citeauthoryear{Hinshaw et~al.,}{Hinshaw
  et~al.}{2013}]{Hinshaw_2013}
Hinshaw G.,  et~al., 2013, \mn@doi [The Astrophysical Journal Supplement
  Series] {10.1088/0067-0049/208/2/19}, 208, 19

\bibitem[\protect\citeauthoryear{Hu \& Dodelson}{Hu \&
  Dodelson}{2002}]{Hu_2002}
Hu W.,  Dodelson S.,  2002, \mn@doi [Annual Review of Astronomy and
  Astrophysics] {10.1146/annurev.astro.40.060401.093926}, 40, 171–216

\bibitem[\protect\citeauthoryear{Huterer et~al.,}{Huterer
  et~al.}{2015}]{Huterer_2015}
Huterer D.,  et~al., 2015, \mn@doi [Astroparticle Physics]
  {10.1016/j.astropartphys.2014.07.004}, 63, 23–41

\bibitem[\protect\citeauthoryear{Ishak}{Ishak}{2018}]{Ishak_2018}
Ishak M.,  2018, \mn@doi [Living Reviews in Relativity]
  {10.1007/s41114-018-0017-4}, 22

\bibitem[\protect\citeauthoryear{Ivezić et~al.}{Ivezić
  et~al.}{2019}]{Ivezi__2019}
Ivezić et~al., 2019, \mn@doi [The Astrophysical Journal]
  {10.3847/1538-4357/ab042c}, 873, 111

\bibitem[\protect\citeauthoryear{Joyce, Lombriser  \& Schmidt}{Joyce
  et~al.}{2016}]{Joyce_2016}
Joyce A.,  Lombriser L.,   Schmidt F.,  2016, \mn@doi [Annual Review of Nuclear
  and Particle Science] {10.1146/annurev-nucl-102115-044553}, 66, 95–122

\bibitem[\protect\citeauthoryear{Kaiser \& Sfakianakis}{Kaiser \&
  Sfakianakis}{2014}]{Kaiser_2014}
Kaiser D.~I.,  Sfakianakis E.~I.,  2014, \mn@doi [Physical Review Letters]
  {10.1103/physrevlett.112.011302}, 112

\bibitem[\protect\citeauthoryear{Kallosh \& Linde}{Kallosh \&
  Linde}{2013}]{Kallosh_2013}
Kallosh R.,  Linde A.,  2013, \mn@doi [Journal of Cosmology and Astroparticle
  Physics] {10.1088/1475-7516/2013/07/002}, 2013, 002–002

\bibitem[\protect\citeauthoryear{Kallosh, Linde  \& Roest}{Kallosh
  et~al.}{2013}]{Kallosh_2013alpha}
Kallosh R.,  Linde A.,   Roest D.,  2013, \mn@doi [Journal of High Energy
  Physics] {10.1007/jhep11(2013)198}, 2013

\bibitem[\protect\citeauthoryear{Kallosh, Linde  \& Roest}{Kallosh
  et~al.}{2014}]{Kallosh_2014}
Kallosh R.,  Linde A.,   Roest D.,  2014, \mn@doi [Journal of High Energy
  Physics] {10.1007/jhep08(2014)052}, 2014

\bibitem[\protect\citeauthoryear{Kazin et~al.,}{Kazin
  et~al.}{2014}]{Kazin_2014}
Kazin E.~A.,  et~al., 2014, \mn@doi [Monthly Notices of the Royal Astronomical
  Society] {10.1093/mnras/stu778}, 441, 3524–3542

\bibitem[\protect\citeauthoryear{{Kodama} \& {Sasaki}}{{Kodama} \&
  {Sasaki}}{1984}]{1984PThPS..78....1K}
{Kodama} H.,  {Sasaki} M.,  1984, \mn@doi [Progress of Theoretical Physics
  Supplement] {10.1143/PTPS.78.1}, \href
  {https://ui.adsabs.harvard.edu/abs/1984PThPS..78....1K} {78, 1}

\bibitem[\protect\citeauthoryear{Kofman, Linde  \& Starobinsky}{Kofman
  et~al.}{1985}]{Kofman:1985aw}
Kofman L.~A.,  Linde A.~D.,   Starobinsky A.~A.,  1985, \mn@doi [Phys. Lett. B]
  {10.1016/0370-2693(85)90381-8}, 157, 361

\bibitem[\protect\citeauthoryear{Lesgourgues}{Lesgourgues}{2011}]{lesgourgues2011cosmic}
Lesgourgues J.,  2011, The Cosmic Linear Anisotropy Solving System (CLASS) I:
  Overview (\mn@eprint {arXiv} {1104.2932})

\bibitem[\protect\citeauthoryear{Lewis}{Lewis}{2019}]{lewis2019getdist}
Lewis A.,  2019, GetDist: a Python package for analysing Monte Carlo samples
  (\mn@eprint {arXiv} {1910.13970})

\bibitem[\protect\citeauthoryear{Li, Li, Wang  \& Wang}{Li
  et~al.}{2011}]{Li_2011}
Li M.,  Li X.-D.,  Wang S.,   Wang Y.,  2011, \mn@doi [Communications in
  Theoretical Physics] {10.1088/0253-6102/56/3/24}, 56, 525–604

\bibitem[\protect\citeauthoryear{Linde}{Linde}{1982}]{Linde:1981mu}
Linde A.~D.,  1982, \mn@doi [Phys. Lett. B] {10.1016/0370-2693(82)91219-9},
  108, 389

\bibitem[\protect\citeauthoryear{Linde}{Linde}{1983}]{Linde:1983gd}
Linde A.~D.,  1983, \mn@doi [Phys. Lett. B] {10.1016/0370-2693(83)90837-7},
  129, 177

\bibitem[\protect\citeauthoryear{Linde}{Linde}{2015}]{Linde_2015}
Linde A.,  2015, \mn@doi [Journal of Cosmology and Astroparticle Physics]
  {10.1088/1475-7516/2015/02/030}, 2015, 030–030

\bibitem[\protect\citeauthoryear{Linde, Noorbala  \& Westphal}{Linde
  et~al.}{2011}]{Linde_2011}
Linde A.,  Noorbala M.,   Westphal A.,  2011, \mn@doi [Journal of Cosmology and
  Astroparticle Physics] {10.1088/1475-7516/2011/03/013}, 2011, 013–013

\bibitem[\protect\citeauthoryear{Linder}{Linder}{2007}]{Linder_2007}
Linder E.~V.,  2007, \mn@doi [General Relativity and Gravitation]
  {10.1007/s10714-007-0550-z}, 40, 329–356

\bibitem[\protect\citeauthoryear{Lobo}{Lobo}{2008}]{lobo2008dark}
Lobo F. S.~N.,  2008, The dark side of gravity: Modified theories of gravity
  (\mn@eprint {arXiv} {0807.1640})

\bibitem[\protect\citeauthoryear{Ludwick}{Ludwick}{2018}]{PhysRevD.98.043519}
Ludwick K.~J.,  2018, \mn@doi [Phys. Rev. D] {10.1103/PhysRevD.98.043519}, 98,
  043519

\bibitem[\protect\citeauthoryear{L’Huillier, Shafieloo, Hazra, Smoot  \&
  Starobinsky}{L’Huillier et~al.}{2018}]{L_Huillier_2018}
L’Huillier B.,  Shafieloo A.,  Hazra D.~K.,  Smoot G.~F.,   Starobinsky
  A.~A.,  2018, \mn@doi [Monthly Notices of the Royal Astronomical Society]
  {10.1093/mnras/sty745}, 477, 2503–2512

\bibitem[\protect\citeauthoryear{{Ma} \& {Bertschinger}}{{Ma} \&
  {Bertschinger}}{1995}]{1995ApJ...455....7M}
{Ma} C.-P.,  {Bertschinger} E.,  1995, \mn@doi [\apj] {10.1086/176550}, \href
  {https://ui.adsabs.harvard.edu/abs/1995ApJ...455....7M} {455, 7}

\bibitem[\protect\citeauthoryear{Malik, Wands  \& Ungarelli}{Malik
  et~al.}{2003}]{PhysRevD.67.063516}
Malik K.~A.,  Wands D.,   Ungarelli C.,  2003, \mn@doi [Phys. Rev. D]
  {10.1103/PhysRevD.67.063516}, 67, 063516

\bibitem[\protect\citeauthoryear{McAllister, Silverstein  \&
  Westphal}{McAllister et~al.}{2010}]{McAllister_2010}
McAllister L.,  Silverstein E.,   Westphal A.,  2010, \mn@doi [Physical Review
  D] {10.1103/physrevd.82.046003}, 82

\bibitem[\protect\citeauthoryear{Miranda, Fabris  \& Piattella}{Miranda
  et~al.}{2017}]{Miranda_2017}
Miranda T.,  Fabris J.~C.,   Piattella O.~F.,  2017, \mn@doi [Journal of
  Cosmology and Astroparticle Physics] {10.1088/1475-7516/2017/09/041}, 2017,
  041–041

\bibitem[\protect\citeauthoryear{Mukhanov \& Chibisov}{Mukhanov \&
  Chibisov}{1981}]{Mukhanov:1981xt}
Mukhanov V.~F.,  Chibisov G.~V.,  1981, JETP Lett., 33, 532

\bibitem[\protect\citeauthoryear{{Mukhanov}, {Feldman}  \&
  {Brandenberger}}{{Mukhanov} et~al.}{1992}]{1992PhR...215..203M}
{Mukhanov} V.~F.,  {Feldman} H.~A.,   {Brandenberger} R.~H.,  1992, \mn@doi
  [\physrep] {10.1016/0370-1573(92)90044-Z}, \href
  {https://ui.adsabs.harvard.edu/abs/1992PhR...215..203M} {215, 203}

\bibitem[\protect\citeauthoryear{Nesseris \& Perivolaropoulos}{Nesseris \&
  Perivolaropoulos}{2008}]{PhysRevD.77.023504}
Nesseris S.,  Perivolaropoulos L.,  2008, \mn@doi [Phys. Rev. D]
  {10.1103/PhysRevD.77.023504}, 77, 023504

\bibitem[\protect\citeauthoryear{Nojiri, Odintsov  \& Tsujikawa}{Nojiri
  et~al.}{2005}]{Nojiri_2005}
Nojiri S.,  Odintsov S.~D.,   Tsujikawa S.,  2005, \mn@doi [Physical Review D]
  {10.1103/physrevd.71.063004}, 71

\bibitem[\protect\citeauthoryear{Padmanabhan}{Padmanabhan}{2002}]{Padmanabhan_2002}
Padmanabhan T.,  2002, \mn@doi [Physical Review D]
  {10.1103/physrevd.66.021301}, 66

\bibitem[\protect\citeauthoryear{Parkinson et~al.,}{Parkinson
  et~al.}{2012}]{Parkinson_2012}
Parkinson D.,  et~al., 2012, \mn@doi [Physical Review D]
  {10.1103/physrevd.86.103518}, 86

\bibitem[\protect\citeauthoryear{Percival \& White}{Percival \&
  White}{2009}]{Percival_2009}
Percival W.~J.,  White M.,  2009, \mn@doi [Monthly Notices of the Royal
  Astronomical Society] {10.1111/j.1365-2966.2008.14211.x}, 393, 297–308

\bibitem[\protect\citeauthoryear{Perlmutter et~al.,}{Perlmutter
  et~al.}{1999}]{Perlmutter_1999}
Perlmutter S.,  et~al., 1999, \mn@doi [The Astrophysical Journal]
  {10.1086/307221}, 517, 565–586

\bibitem[\protect\citeauthoryear{Piazza \& Tsujikawa}{Piazza \&
  Tsujikawa}{2004}]{Piazza_2004}
Piazza F.,  Tsujikawa S.,  2004, \mn@doi [Journal of Cosmology and
  Astroparticle Physics] {10.1088/1475-7516/2004/07/004}, 2004, 004–004

\bibitem[\protect\citeauthoryear{{Planck Collaboration}, {Aghanim}
  et~al.}{{Planck Collaboration} et~al.}{2020}]{2020A&A...641A...5P}
{Planck Collaboration} {Aghanim} N.,   et~al., 2020, \mn@doi [\aap]
  {10.1051/0004-6361/201936386}, \href
  {https://ui.adsabs.harvard.edu/abs/2020A&A...641A...5P} {641, A5}

\bibitem[\protect\citeauthoryear{Ratra \& Peebles}{Ratra \&
  Peebles}{1988}]{PhysRevD.37.3406}
Ratra B.,  Peebles P. J.~E.,  1988, \mn@doi [Phys. Rev. D]
  {10.1103/PhysRevD.37.3406}, 37, 3406

\bibitem[\protect\citeauthoryear{Riess et~al.,}{Riess
  et~al.}{1998}]{Riess_1998}
Riess A.~G.,  et~al., 1998, \mn@doi [The Astronomical Journal]
  {10.1086/300499}, 116, 1009–1038

\bibitem[\protect\citeauthoryear{{Ross}, {Samushia}, {Howlett}, {Percival},
  {Burden}  \& {Manera}}{{Ross} et~al.}{2015}]{2015MNRAS.449..835R}
{Ross} A.~J.,  {Samushia} L.,  {Howlett} C.,  {Percival} W.~J.,  {Burden} A.,
  {Manera} M.,  2015, \mn@doi [\mnras] {10.1093/mnras/stv154}, \href
  {https://ui.adsabs.harvard.edu/abs/2015MNRAS.449..835R} {449, 835}

\bibitem[\protect\citeauthoryear{Sahni}{Sahni}{2002}]{Sahni_2002}
Sahni V.,  2002, \mn@doi [Classical and Quantum Gravity]
  {10.1088/0264-9381/19/13/304}, 19, 3435–3448

\bibitem[\protect\citeauthoryear{Salopek, Bond  \& Bardeen}{Salopek
  et~al.}{1989}]{PhysRevD.40.1753}
Salopek D.~S.,  Bond J.~R.,   Bardeen J.~M.,  1989, \mn@doi [Phys. Rev. D]
  {10.1103/PhysRevD.40.1753}, 40, 1753

\bibitem[\protect\citeauthoryear{{Scolnic} et~al.}{{Scolnic}
  et~al.}{2018}]{2018ApJ...859..101S}
{Scolnic} D.~M.,  et~al., 2018, \mn@doi [\apj] {10.3847/1538-4357/aab9bb},
  \href {https://ui.adsabs.harvard.edu/abs/2018ApJ...859..101S} {859, 101}

\bibitem[\protect\citeauthoryear{Shahalam, Myrzakulov, Myrzakul  \&
  Wang}{Shahalam et~al.}{2018}]{Shahalam_2018}
Shahalam M.,  Myrzakulov R.,  Myrzakul S.,   Wang A.,  2018, \mn@doi
  [International Journal of Modern Physics D] {10.1142/s021827181850058x}, 27,
  1850058

\bibitem[\protect\citeauthoryear{Silverstein \& Westphal}{Silverstein \&
  Westphal}{2008}]{PhysRevD.78.106003}
Silverstein E.,  Westphal A.,  2008, \mn@doi [Phys. Rev. D]
  {10.1103/PhysRevD.78.106003}, 78, 106003

\bibitem[\protect\citeauthoryear{Song \& Percival}{Song \&
  Percival}{2009}]{Song_2009}
Song Y.-S.,  Percival W.~J.,  2009, \mn@doi [Journal of Cosmology and
  Astroparticle Physics] {10.1088/1475-7516/2009/10/004}, 2009, 004–004

\bibitem[\protect\citeauthoryear{Spergel et~al.,}{Spergel
  et~al.}{2015}]{spergel2015widefield}
Spergel D.,  et~al., 2015, Wide-Field InfrarRed Survey Telescope-Astrophysics
  Focused Telescope Assets WFIRST-AFTA 2015 Report (\mn@eprint {arXiv}
  {1503.03757})

\bibitem[\protect\citeauthoryear{Starobinsky}{Starobinsky}{1980}]{Starobinsky:1980te}
Starobinsky A.~A.,  1980, \mn@doi [Phys. Lett. B]
  {10.1016/0370-2693(80)90670-X}, 91, 99

\bibitem[\protect\citeauthoryear{Starobinsky}{Starobinsky}{1983}]{Starobinsky:1983zz}
Starobinsky A.~A.,  1983, Sov. Astron. Lett., 9, 302

\bibitem[\protect\citeauthoryear{Steinhardt}{Steinhardt}{2003}]{Steinhardt:2003st}
Steinhardt P.~J.,  2003, \mn@doi [Phil. Trans. Roy. Soc. Lond. A]
  {10.1098/rsta.2003.1290}, 361, 2497

\bibitem[\protect\citeauthoryear{Torrado \& Lewis}{Torrado \&
  Lewis}{2021}]{Torrado_2021}
Torrado J.,  Lewis A.,  2021, \mn@doi [Journal of Cosmology and Astroparticle
  Physics] {10.1088/1475-7516/2021/05/057}, 2021, 057

\bibitem[\protect\citeauthoryear{Tsujikawa}{Tsujikawa}{2011}]{Tsujikawa_2011}
Tsujikawa S.,  2011, \mn@doi [Astrophysics and Space Science Library]
  {10.1007/978-90-481-8685-3_8}, p. 331–402

\bibitem[\protect\citeauthoryear{Tsujikawa}{Tsujikawa}{2013}]{Tsujikawa_2013}
Tsujikawa S.,  2013, \mn@doi [Classical and Quantum Gravity]
  {10.1088/0264-9381/30/21/214003}, 30, 214003

\bibitem[\protect\citeauthoryear{Velten, vom Marttens  \& Zimdahl}{Velten
  et~al.}{2014}]{Velten_2014}
Velten H. E.~S.,  vom Marttens R.~F.,   Zimdahl W.,  2014, \mn@doi [The
  European Physical Journal C] {10.1140/epjc/s10052-014-3160-4}, 74

\bibitem[\protect\citeauthoryear{Väliviita, Majerotto  \& Maartens}{Väliviita
  et~al.}{2008}]{V_liviita_2008}
Väliviita J.,  Majerotto E.,   Maartens R.,  2008, \mn@doi [Journal of
  Cosmology and Astroparticle Physics] {10.1088/1475-7516/2008/07/020}, 2008,
  020

\bibitem[\protect\citeauthoryear{Whitt}{Whitt}{1984}]{Whitt:1984pd}
Whitt B.,  1984, \mn@doi [Phys. Lett. B] {10.1016/0370-2693(84)90332-0}, 145,
  176

\makeatother
\end{thebibliography}

%\begin{thebibliography}{99}
%\end{thebibliography}

%%%%%%%%%%%%%%%%%%%%%%%%%%%%%%%%%%%%%%%%%%%%%%%%%%

% Don't change these lines
\bsp	% typesetting comment
\label{lastpage}
\end{document}